\documentclass[fleqn,usenatbib]{mnras}

\usepackage{newtxtext,newtxmath}

\usepackage[T1]{fontenc}
\usepackage{ae,aecompl}
\usepackage{natbib}
\bibpunct{(}{)}{;}{a}{}{,}

\usepackage{graphicx}	
\usepackage{amsmath}	
\usepackage{amssymb}	
\usepackage{txfonts}
\usepackage{textcomp}

\def\HI{H{\,\small I}}

\defcitealias{Heywood13}{Heywood et al. (2013}
\defcitealias{Windhorst90}{Windhorst et al. (1990}

\title[A 1.4 GHz mosaic of the Lockman Hole region]{The Lockman Hole Project: New constraints on  the sub-mJy source counts from a wide-area 1.4~GHz mosaic }

\author[I. Prandoni et al.]{
I. Prandoni$^{1}$\thanks{E-mail: prandoni@ira.inaf.it (IP)},
G. Guglielmino$^{1,2}$,
R. Morganti$^{3,4}$,
M. Vaccari$^{5,1}$, 
A. Maini$^{1,2,6,7}$,
\newauthor H.~J.~A. R\"ottgering$^{8}$, M.~J. Jarvis$^{9,5}$ and M.A. Garrett$^{10,11}$ 
\\
$^{1}$INAF-Istituto di Radioastronomia, via Gobetti 101, Bologna 40129, Italy\\
$^{2}$University of Bologna, DIFA, Via Ranzani 2, Bologna 40126, Italy\\
$^{3}$ASTRON, the Netherlands Institute of Radio Astronomy, Postbus 2, 7990 AA, Dwingeloo, The Netherlands\\
$^{4}$Kapteyn Astronomical Institute, University of Groningen, P.O. Box 800,9700 AV Groningen, The Netherlands\\
$^{5}$Department of Physics \& Astronomy, University of the Western Cape, Bellville 7535, Cape Town, South Africa\\
$^{6}$Department of Physics and Astronomy, Macquarie University, Balaclava Road, North Ryde, 2109 NSW, Australia\\
$^{7}$CSIRO Astronomy \& Space Science, PO Box 76, Epping, NSW 1710, Australia\\
$^{8}$Leiden Observatory, Leiden University, P.O. Box 9513, NL-2300 RA Leiden, The Netherlands\\
$^{9}$Oxford Astrophysics, Denys Wilkinson Building, Keble Road, Oxford OX1 3RH, UK\\
$^{10}$The Alan Turing Building, School of Physics and Astronomy, Oxford Road, University of Manchester, M13 9PL, UK\\
$^{11}$Jodrell Bank Observatory, Lower Withington, University of Manchester, Macclesfield,  Cheshire, SK11 9DL, UK
}

\date{Accepted XXX. Received YYY; in original form ZZZ}

\pubyear{2016}

\begin{document}
\label{firstpage}
\pagerange{\pageref{firstpage}--\pageref{lastpage}}
\maketitle

\begin{abstract}
This paper is part of a series discussing the results obtained in the framework of a wide international collaboration - the Lockman Hole Project - aimed at improving the extensive multi-band coverage available in the Lockman Hole region, through novel deep, wide-area, multi-frequency (60, 150, 350 MHz and 1.4 GHz) radio surveys.  This multi-frequency, multi-band information will be exploited to get a comprehensive view of star formation and AGN activities in the high redshift Universe from a radio perspective.  In this paper we present  novel 1.4 GHz mosaic observations 
obtained with the  Westerbork Synthesis Radio Telescope (WSRT). With an area coverage of 6.6 square degrees, this is the largest survey reaching an rms noise of 11 $\mu$Jy/b.  In this paper we present the source catalogue ($\sim 6000$ sources with flux densities $S\ga 55$ $\mu$Jy (5$\sigma$), and we discuss the 1.4 GHz source counts derived from it. Our source counts provide very robust statistics in the flux range $0.1<\rm{S}<1$ mJy, and are in excellent agreement with other robust determinations obtained at lower and higher flux densities.  A clear excess is found with respect to the counts predicted by the semi-empirical radio sky simulations developed in the framework of  the SKA Simulated Skies  project. A  preliminary analysis of the identified (and classified) sources  suggests this excess is to be ascribed to star forming galaxies, which seem to show a steeper evolution than predicted. 
\end{abstract}




\section{Introduction}
\label{sec:intro}

After many years of extensive multi-band follow-up studies  it is now established that the sub-mJy population has a composite nature. Radio-loud (RL)  Active Galactic Nuclei (AGN) remain largely dominant down to flux densities of $400-500$ $\mu$Jy (e.g. \citealt{Mignano08}), while star-forming galaxies (SFG) become the dominant population below  $\sim 100$ $\mu$Jy (e.g. \citealt{Simpson06,Seymour08,Smolcic08}). More recently it has been shown that a significant fraction of the sources below 100 $\mu$Jy show signatures of AGN activity at non-radio wavelengths (e.g. Seyfert galaxies or QSO).  These AGNs are often referred to in the literature as
radio-quiet (RQ) AGN (see e.g. \citealt{Padovani09,Padovani11,Padovani15,Bonzini13}), because the vast majority of them do not display large scale jets or lobes. It is worth noting that these systems are  typically radiatively  efficient AGNs, characterized by high accretion rates ($\ga 1\%$), while the low-luminosity RL AGN population detected at sub-mJy fluxes  is largely made of systems hosted by early-type galaxies (\citealt{Mignano08}), likely characterized by radiatively inefficient, low accretion rates ($<<1\%$). In other words a classification based on radio loudness (despite not being fully appropriate for faint radio-selected AGNs\footnote{A detailed discussion of AGN classification in view of the latest results from deep radio surveys, is presented in \citet{Padovani17}, who proposes to update the terms RL/RQ AGNs into jetted/non-jetted AGNs, based on the presence/lack of strong relativistic jets.}) implies, at least in a statistical sense,  a more profound distinction between fundamental AGN classes (for a comprehensive review on AGN types and properties we refer to \citealt{Heckman14}).  

The presence of large numbers of AGN-related sources at sub-mJy/$\mu$Jy radio flux densities has given a new interesting scientific perspective to deep radio surveys, as they provide a powerful dust/gas-obscuration-free tool to get a global census of both star formation and AGN activity (and related AGN feedback) up to very high redshift and down to the radio-quiet AGN regime (see \citealt{Padovani16} for a comprehensive review).  However several   uncertainties remain due to observational issues and limitations. First the radio source counts show a large scatter below $\sim 1$ mJy, resulting in a large uncertainty on the actual  radio source number density at sub-mJy flux density levels. This scatter can be largely ascribed to cosmic variance effects \citep{Heywood13}, but may also be due, at least in some cases, to survey systematics (see e.g. \citealt{Condon12}, and discussion in Sect.~\ref{sec-counts}). Secondly, for a full and robust characterization of the faint radio population the availability of deep multi-wavelength ancillary datasets is essential, but typically limited to very small regions of the sky.
Mid- and far-infrared (IR) data, as well as deep X-ray information, for example, has proved to be crucial to reliably separate SFGs from RQ AGNs (see e.g. \citealt{Bonzini13,Bonzini15}).  When available, optical/near-IR spectroscopy is of extreme value, as it provides source redshifts and, if of sufficient quality, a very reliable  classification of the host galaxies (SFGs, Seyferts, QSO, etc.), through the analysis of line profiles (broad vs narrow) and line ratios (see e.g. the diagnostic diagrams introduced  by \citealt{Baldwin81} and later revised by \citealt{Veilleux87}). Alternatively, multi-band optical/IR photometry can be used: host galaxies can be classified through their colors and/or Spectral Energy Distributions (SED), and stellar masses and photometric estimates of the source redshifts can be derived (several statistical methods and tools are presented in the literature; for a recent application to deep radio continuum surveys, see 
\citealt{Duncan18a,Duncan18b}). 

Finally, the origin of the radio emission in RQ AGNs is currently hotly debated. Most radio-selected RQ AGNs are characterized by compact sizes, i.e. they are unresolved or barely resolved at a few arcsec scale, which is similar to the host galaxy size. RQ AGNs have also been found to share properties with SFGs. They have similar radio spectra and luminosities \citep{Bonzini13,Bonzini15}; their radio luminosity functions show similar evolutionary trends \citep{Padovani11}; their host galaxies have similar colours, optical morphologies, and stellar masses \citep{Bonzini13}. For all these reasons it was concluded that the radio emission in  RQ AGNs is triggered by star formation \citep{Padovani11,Bonzini13,Bonzini15,Ocran17}. On the other hand high-resolution radio follow ups of RQ AGN samples with Very Long Baseline Interferometry (VLBI) arrays have shown that  a significant fraction of RQ AGNs  (20-40\%, depending on the sample) contain AGN cores  that contribute significantly (50\% or more) to the total radio emission \citep{Maini16,Herrera16,Herrera17}. A different approach was followed by \citet{Delvecchio17} in the framework of the VLA COSMOS 3~GHz Project.  To identify possible AGN contributions, they first exploited the dense multi-band information in the COSMOS field to derive accurate star formation rates (SFR) via SED fitting; then they analyzed the ratio between the 1.4 GHz radio luminosity and the SFR for each source.  This resulted in $\sim 30\%$ of the sources with AGN signatures at non-radio wavelengths displaying a significant ($>3\sigma$) radio excess.  It is worth noticing that radio selection does not seem to play a major role here. Controversial results arise also from investigations of optically-selected QSOs, with authors claiming a pure star formation origin of their radio emission \citep{Kimball11,Condon13}, and others providing evidence of the presence of a radio luminosity excess with respect to star forming galaxies of similar masses \citep{White15}. Such an excess appears to be correlated with the optical luminosity \citep{White17}. \\
The most likely scenario is  that RQ AGN are composite systems where star formation and AGN triggered radio emission can coexist, over a wide range of relative contributions. This scenario is  supported by the recent modeling work of \citet{Mancuso17}, who showed that the observed radio counts can be very well reproduced by a three-component population (SFGs, RL and RQ AGN), where RQ AGN are the sum of two sub-components: one dominated by star formation (so-called radio silent), and the other by AGN-triggered radio emission. 

In order to overcome the aforementioned issues about cosmic variance and limited multi-band information, 
deep radio samples over  wide areas ($>>1$ sq. degr.) are needed, in regions where 
wide-area, deep multi-band ancillary data are available. This is a pre-requisite to get robust estimations of the sub-mJy radio source number density and of the fractional contribution of each class of sources as a function of cosmic time in representative volumes of the Universe (i.e. not biased by cosmic variance). At the same time wide-area surveys allow us to probe AGN and/or star formation activities in a variety of different environments.  Additional important information may come from multi-frequency radio coverage: radio spectra may help to constrain the origin of the radio emission in the observed sources and to understand  its link to the host galaxy bolometric emission. This is especially true if high-resolution radio data are available and source structures can be inferred. 
Star-forming galaxies 
typically have a {\it steep} radio spectral index ($\alpha \sim -0.7$/$-0.8$, where $S\propto \nu^\alpha$), with a relatively small dispersion ($\pm 0.24$, \citealt{Condon92}). Radio spectral index studies combined with source structure information (radio jets and lobes) may thus help to disentangle star-forming from steep-spectrum radio galaxy populations. A {\it flat} ($\alpha>-0.5$) radio spectral index can identify  core-dominated AGNs \citep{Blundell07} and GHz-peaked sources (GPS; \citealt*{Gopal83}; \citealt{Odea98,Snellen00}).  
Ultra-steep radio spectra ($\alpha<-1$; \citealt{Rottgering94,Chambers96,Jarvis01}) are a typical feature of high-redshift ($z>>2$) radio galaxies. 

The Lockman Hole (LH, \citealt{Lockman86}) is one of the best studied extra-galactic regions of the sky (see Sect.~\ref{sec-multiband} for a comprehensive summary of the available multi-band coverage in this region). Given its high declination ($\sim$ +58$^{\circ}$), the LH is also best suited for deep, high-resolution, high-fidelity imaging with the LOw-Frequency ARray (LOFAR). 

The Lockman Hole Project is an international collaboration aimed at extending the multi-band information available in the LH region, through novel multi-frequency radio surveys down to 60-150 MHz, a frequency domain that is now accessible for wide-area deep fields thanks to the combination of field of view, sensitivity and spatial resolution of LOFAR. 
This information, together with the available ancillary data, will allow us to get robust observational constraints on the faint extra-galactic radio sky, in preparation for next-generation continuum extra-galactic surveys with ASKAP \citep{Johnston07}, MeerKat \citep{Booth12}, and ultimately the Square Kilometre Array (SKA).

This paper presents a Westerbork (WSRT) 1.4 GHz mosaic covering $\sim 6.6$ deg$^2$ down to 11 $\mu$ beam$^{-1}$ rms and the source catalogue extracted from it. The 345 MHz follow-up, again obtained  with the WSRT, is presented in a following paper (Prandoni et al. in prep.), while  the first LOFAR observations of this region are  presented in \citet{Mahony16}. 

This paper is organized as follows. Section~\ref{sec-multiband} gives an overview of the multi-wavelength data available for the Lockman Hole region. In Sections~\ref{sec-observations}, we describe the WSRT 1.4 GHz observations, the related data reduction and the analysis performed to characterize the noise properties of the final mosaic. In Section~\ref{sec-extraction}  we describe the method used to extract the sources and the final catalogue obtained. In Section~\ref{sec-errors} and \ref{sec-sizes} we provide estimates of the source parameters' errors and we analyze possible systematic effects. In Section~\ref{sec-counts} we present the source counts derived from the present catalogue and we discuss them in comparison with other existing source counts obtained from wide-area 1.4 GHz surveys. In section~\ref{sec-countsid}  we assess the contribution of each class of sources to our overall radio source counts, based on a preliminary analysis of the radio source optical/IR properties (presented in detail in a following paper), and we compare it to existing modeling predictions. In Section~\ref{sec-summary} we summarize our main results.

   \begin{figure*}
   \centering
   \resizebox{1.2\textwidth}{!}{\includegraphics[trim=100 0 0 0]{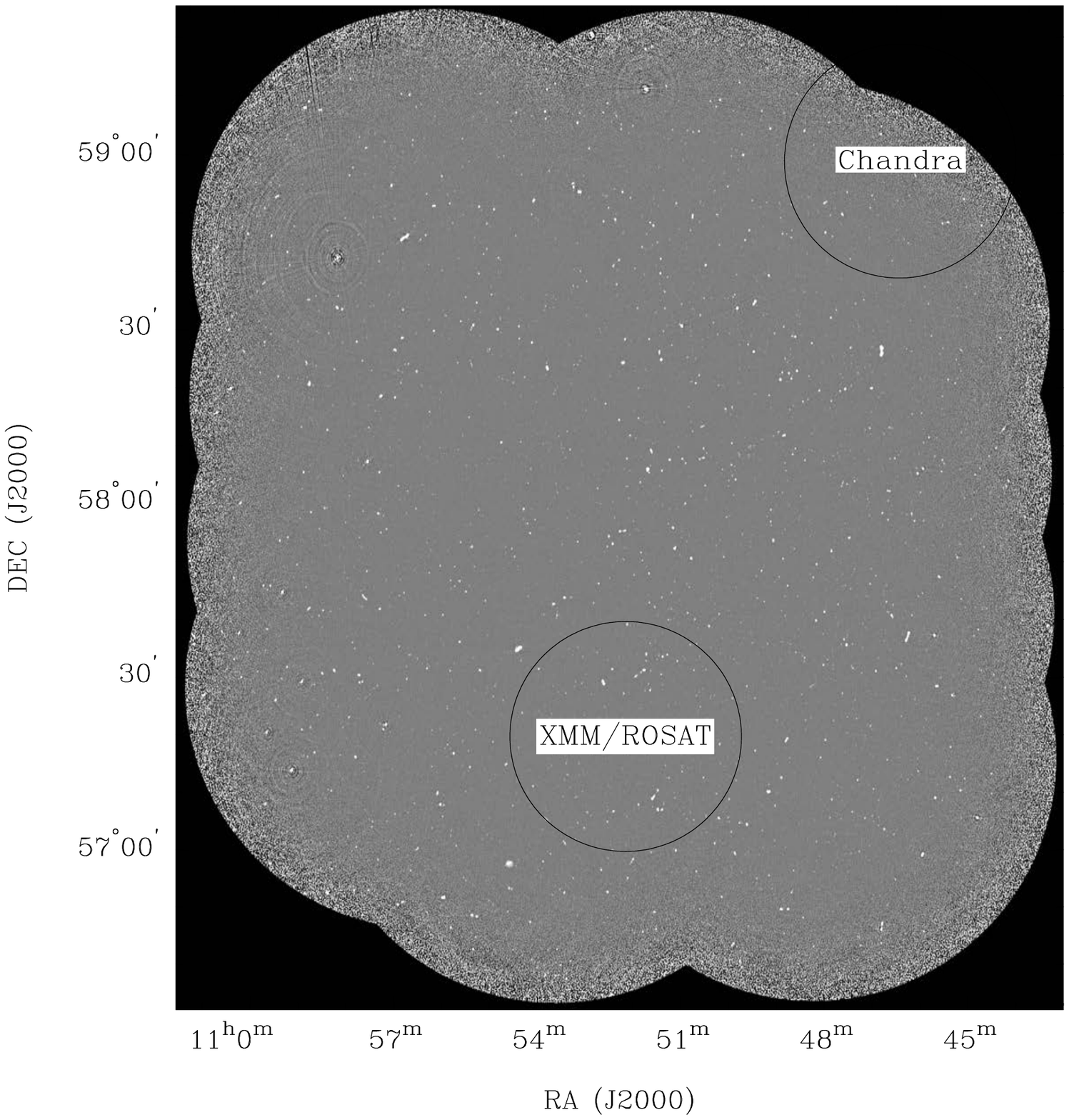}}
      \caption{The WSRT 1.4 GHz mosaic: 16 overlapping pointings, with spacing of $22^{\prime}$ in R.A. and $25^{\prime}$ in Dec. Highlighted are the two locations of the deep X-ray fields, where most existing deep radio observations have been taken (see text for more details).}
         \label{fig-mosaic}
   \end{figure*}

\section{Multi-wavelength coverage of the Lockman Hole Region}
\label{sec-multiband}

The LH is the region of lowest \HI ~column density in the sky. Its low-infrared background (0.38 MJy 
sr$^{-1}$ at 100 $\mu$m; \citealt{Lonsdale03}) makes this region particularly well suited for deep-infrared observations. The Spitzer Space Telescope \citep{Werner04} observed $\sim 12$ deg$^2$ of the LH region in 2004 as part of the Spitzer Wide-area Infrared Extragalactic survey (SWIRE; \citealt{Lonsdale03}). Observations were performed using the Infrared Array Camera (IRAC; \citealt{Fazio04}) operating at 3.6, 4.5, 5.8 and 8 $\mu$m, and the Multiband Imaging Photometer for Spitzer (MIPS; \citealt{Rieke04}) at 24, 70 and 160 $\mu$m. Deeper, confusion-limited observations at 3.6 and 4.5 $\mu$m were obtained over $\sim 4$ deg$^2$ during the warm mission of Spitzer as part of the Spitzer Extragalactic Representative Volume Survey (SERVS, \citealt{Mauduit12}). In addition about 16 deg$^2$ overlapping with the SWIRE survey of the LH have been targeted by the Herschel Space Observatory with the Photoconductor Array Camera and Spectrometer (PACS, 100 and 160 $\mu$m) and the Spectral and Photometric Imaging REceiver (SPIRE, 250, 350  and 500 $\mu$m) as part of the Herschel Multi-tiered Extragalactic Survey (HerMES, \citealt{Oliver12}). 

A great deal of complementary data have been taken on the LH at other wavelengths in order to exploit the availability of sensitive infrared observations, including GALEX GR6Plus7 ultraviolet photometry (Martin et al., 2005), SDSS DR14 optical spectroscopy and photometry in the ugriz bands to a depth of $\sim 22$ mag \citep{Abolfathi18}, INT Wide Field Camera (WFC) optical photometry (u, g, r, i, z down to AB magnitudes 23.9,
24.5, 24.0, 23.3, 22.0 respectively; \citealt{Gonzales11}) and UK Infrared Deep Sky Survey Deep Extragalactic Survey (UKIDSS DXS) DR10Plus photometry in the J and K bands, with a sensitivity of K $\sim 21-21.5$ mag (Vega; \citealt{Lawrence07}).  
There are existing near-infrared data across the region from the Two Micron All Sky Survey (2MASS; \citealt{Beichman03}) to J, H and Ks band magnitudes of 17.8, 16.5 and 16.0.
A photometric redshift catalogue containing 229,238 galaxies and quasars within the LH has been constructed from band-merged data (\citealt{RowanRobinson08}; but see \citealt{RowanRobinson13} for the latest version of the SWIRE photometric redshift catalog, including photometric redshifts and SED models based on optical, near-infrared and Spitzer photometry). Deep surveys within the Lockman Hole region have been undertaken with the Submillimetre Common-User Bolometer Array (SCUBA; \citealt{Holland99}) at 850 $\mu$m \citep{Coppin06}, and with the X-ray satellites ROSAT  \citep{Hasinger98}, XMM-Newton \citep{Hasinger01,Mainieri02,Brunner08} and Chandra \citep{Polletta06}.

A variety of radio surveys cover limited areas within the LH region, in coincidence with the two deep X-ray fields (highlighted in Fig.~\ref{fig-mosaic}). The first of these was by \citet{deruiter97}, who observed
an area of 0.35 deg$^2$ at 1.4 GHz centered on the ROSAT/XMM pointing (R.A.=10:52:09; Dec.=+57:21:34, J2000), using the  Very Large Array (VLA) in C-configuration, with a rms noise level of
$30-55$ $\mu$Jy beam$^{-1}$. A similar deep observation was carried out by \citet{Ciliegi03}, who observed a 0.087 deg$^2$ region at 4.89 GHz using the VLA in C-configuration, with a rms noise level of 11 $\mu$Jy beam$^{-1}$. More recently, \citet{Biggs06} observed  a 320 arcmin$^2$ area, using the VLA at 1.4~GHz
operating in the A- and B-configurations, and with an rms noise level of 4.6 $\mu$Jy beam$^{-1}$. VLA B-configuration 1.4 GHz observations of a larger region (three overlapping VLA pointings) were performed by \citet{Ibar09}, reaching an rms noise of $\sim 6$ $\mu$Jy beam$^{-1}$ in the central 100 arcmin$^2$ area. These observations were matched with 610 MHz GMRT observations down to an rms noise of $\sim 15$ $\mu$Jy beam$^{-1}$. Less sensitive GMRT 610 MHz observations of a much larger area were carried out by \citet{Garn08a,Garn08b}, covering $\sim 5$ deg$^2$ down to an rms noise of $\sim 60$ $\mu$Jy. This survey was later extended to $\sim 13$ deg$^2$ \citep{Garn10}. Two fields of the 10C 15 GHz survey (AMI consortium, 2011; \citealt{Whittam13}) overlap with the LH region, for a total of $\sim 4.6$ deg$^2$. Rms noise levels of $\sim 50-100$ $\mu$Jy where reached at a spatial resolution is $30^{\prime\prime}$.
The deepest 1.4 GHz observations to date (rms noise of $\sim 2.7$ $\mu$Jy beam$^{-1}$ were performed by  \citet{Owen08} at the location of the Chandra deep pointing (R.A.=10:46; Dec.=+59:00, J2000). This was later matched with very sensitive VLA (C-configuration) 324.5 MHz observations down to a rms noise of $\sim 70$ $\mu$Jy beam$^{-1}$ in the central part \citep{Owen09}. This field (also known as Lockman North) has been recently the target of wide-band 3 GHz observations with the upgraded Karl Jansky VLA , reaching an rms noise level of 1.01 $\mu$Jy beam$^{-1}$ \citep{Condon12,Vernstrom14,Vernstrom16a,Vernstrom16b}. The Faint Images of the Radio Sky at Twentycm (FIRST; \citealt{Becker95}) and NRAO VLA Sky Survey (NVSS; \citealt{Condon98}) surveys both cover the entire region at 1.4~GHz, but only to relatively shallow noise levels of 150 and 450 $\mu$Jy beam$^{-1}$, respectively. 

Finally, as part of the Lockman Hole Project, the field has been imaged with WSRT at 350 MHz down to the confusion limit ($\sim 0.5$ mJy rms; Prandoni et al. in prep.), and with LOFAR at 150~MHz ($\sim 160$ $\mu$Jy rms; \citealt{Mahony16}). Deeper 150 MHz LOFAR observations are ongoing (Mandal et al. in prep.).  \citet{Mahony16} present a multi-frequency study of the radio sources in the field, based on most of the afore-mentioned radio observations, including the catalogue presented here. The combination of LOFAR 150~MHz and WSRT 1.4~GHz data, resulted in a sample of 1302 matched sources (see \citealt{Mahony16} for more details). 

\begin{figure*}
  \centering
  \vspace{-2cm}
   \resizebox{9cm}{!}{\includegraphics{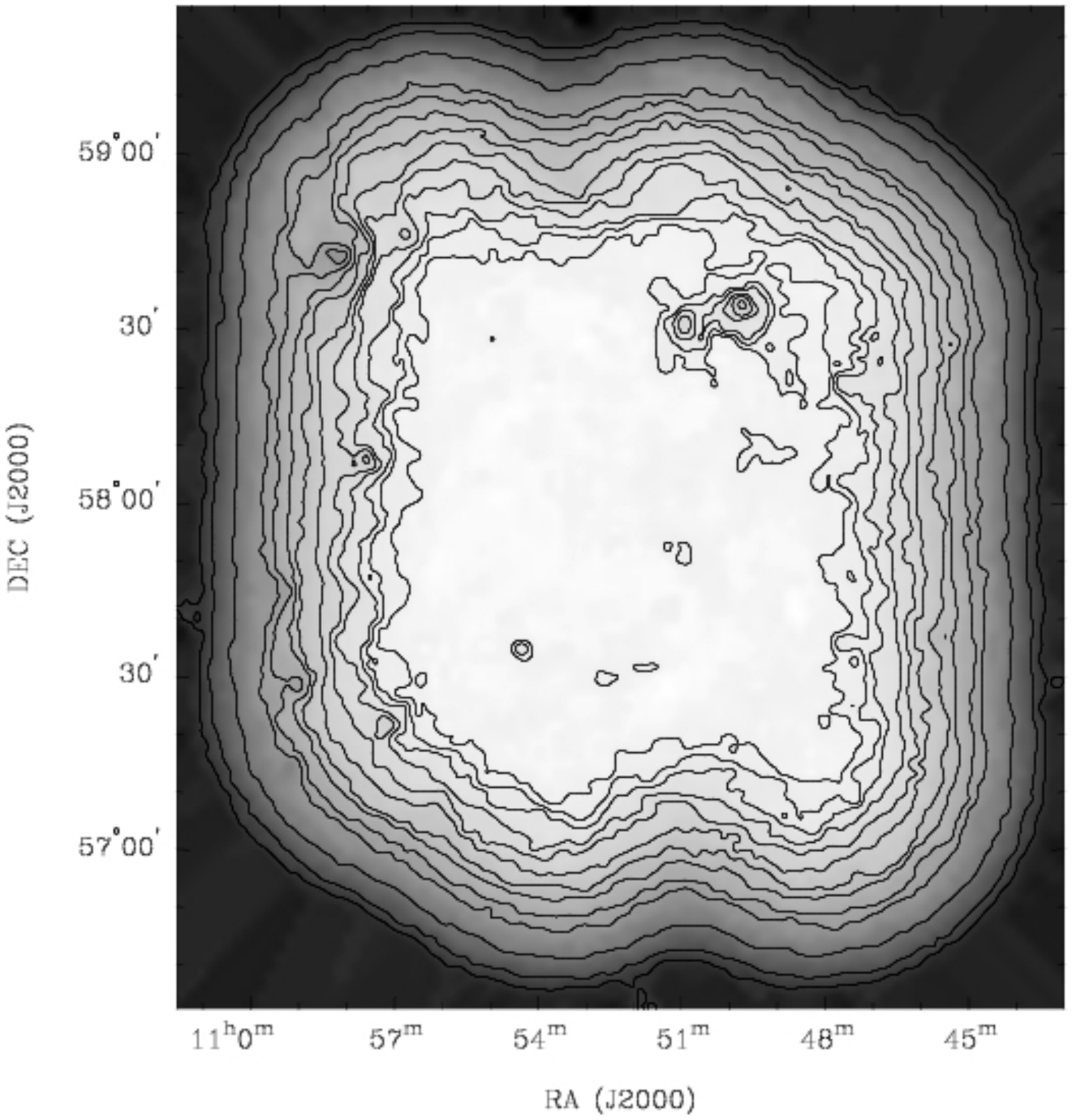}}\resizebox{8.cm}{!}{\includegraphics[trim=0 -200 0 0]{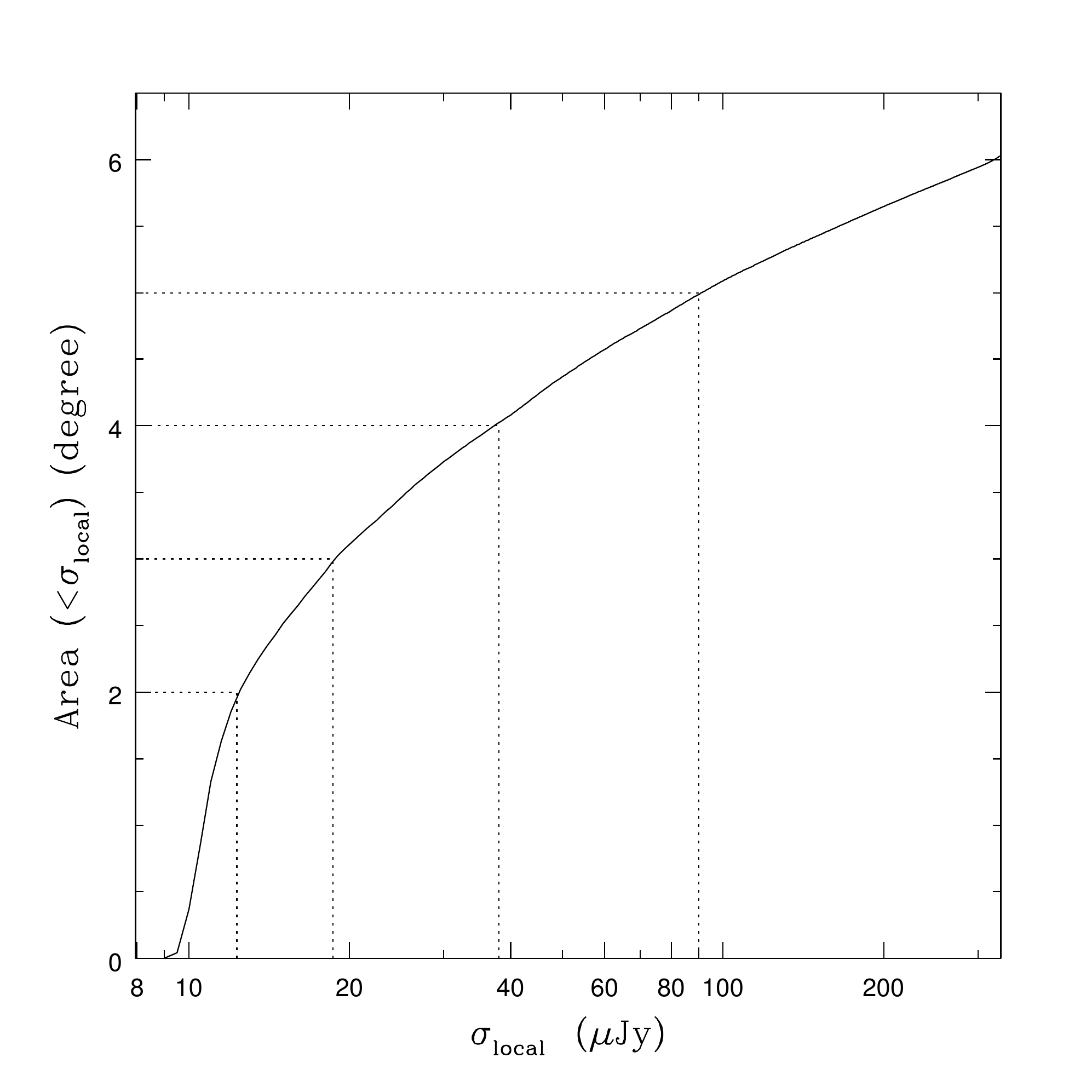}}
  \vspace{-2cm}
   \caption{\emph{Left}: Noise map. Contours refer to 1.1, 1.2, 1.3, 1.5, 1.7, 2, 2.5, 3, 4, 5, 10, 20, 30 multiples of the noise center value (11 $\mu$Jy). \emph{Right}: Visibility area of the WSRT 1.4 GHz mosaic. Cumulative fraction of the total area of the noise map characterized by a measured noise lower than a given value. Dotted lines indicate the maximum noise value measured over  40\%, 60\% and 80\% of the total area.}
              \label{fig-noise}
    \end{figure*}

\section{The new 1.4 GH\lowercase{z} Mosaic}
 \label{sec-observations}
We observed the LH region  with the WSRT at 1.4 GHz, in the period Dec 2006 - Jun 2007. The observations covered an area of $\sim$6.6 square degrees (mostly overlapping the Spitzer and Herschel surveys), through overlapping pointings. A good compromise between uniform sensitivity and observing efficiency  
is generally obtained with a mosaic pattern where pointing spacings, $s$, are $\approx$FWHP', where FWHP'=FWHP/$\sqrt2$, and FWHP is the full width at half power of the primary beam (see \citealt{Prandoni00a}). For our particular case FWHP$\sim$36$^{\prime}$, and FWHP'$\sim25.46^{\prime}$. From noise simulations we got 5\% noise variations with $s$=0.85~FWHP' (=22$^{\prime}$) and 10\% variations with $s$=FWHP'. We then decided to cover the 6.6 deg$^2$ area with 16 overlapping pointings, with spacing of $22^{\prime}$ in RA and $25^{\prime}$ in DEC. Each field was observed for 12h. The primary calibrator (3C48) was observed for 15min at the beginning of each 12h run, and the secondary calibrator (J1035+5628), unresolved on VLBA scale, was observed for 3min every hour. The data was recorded in 512 channels, organized in eight 20 MHz sub-bands, 64 channels each. The channel width is 312 KHz, and the total bandwidth is 160 MHz. 

For the data reduction we used the \emph{Multichannel Image Reconstruction, Image Analysis and Display} (MIRIAD) software package \citep{Sault09}. Each field was calibrated and imaged separately. Imaging and deconvolution was performed in multi-frequency synthesis mode, taking into proper account the spectral variation of the dirty beam over the image during the cleaning process (MIRIAD task MFCLEAN).  Each field was cleaned to a distance of 50 arcmin from the phase center (i.e. down to about the zero point primary beam width) in order to deconvolve all the sources in the field. 
All the images were produced using uniform weighting to get the maximum spatial resolution. Subsequently we combined together all the images to create a single primary beam corrected mosaic (pixel size $=2^{\prime\prime}$). The synthesized beam is  $11^{\prime\prime}\times 9^{\prime\prime}$, with position angle PA=$0^{\circ}$. The resulting 1.4 GHz mosaic, centered at R.A. =10:52:16.6; Dec. = +58:01:15 (J2000), is shown in Fig.~\ref{fig-mosaic}.

\subsection{Noise Map}
\label{sec-noise}

To investigate the noise characteristics of our 1.4~GHz image we constructed a noise map with the software SExtractor \citep{Bertin96}. Although SExtractor was originally developed for the analysis of optical data, it is widely used for noise analysis of radio images as well (see e.g. \citealt{Bondi03,Huynh05, Prandoni06}). SExtractor initially estimates the local background in each mesh from the pixel data. Then the local background histogram is clipped iteratively until convergence is reached at $\pm3\sigma$ around its median. The choice of mesh size is very important. When it is too small, the background tends to be overestimated due to the presence of real sources. When it is too large, any small scale variation of the background is washed out.  A mesh size of $50\times 50$ pixel (approximately $10\times 10$ beams), was found to be appropriate for our case (see also discussion in Sect.~\ref{sec-extraction}). However it should be noted that border effects make the determination of the local noise less reliable in the outermost regions of the mosaic. 

The obtained noise map is shown in Figure~\ref{fig-noise} (left panel). The rms was found to be approximately uniform (noise variations $<10\%$) over the central region, with a value of about 11 $\mu$Jy. Then it radially increases up to $\sim500$ $\mu$Jy at the very border of the mosaic. This is in agreement with the expectations, as better discussed in Section~\ref{sec-extraction}. Sub-regions characterized by noise values higher than the expected ones are found to correspond to very bright sources, due to dynamic range limits introduced by residual phase errors. In Figure~\ref{fig-noise}  (right panel) the total area of the noise map characterized by noise measurements lower than a given value is plotted. The inner $\sim2$ deg$^2$ region is characterized by a noise increment $\leq10\%$ (noise values $\leq 12$ $\mu$Jy). Noise increments to 18, 38 and 90 $\mu$Jy are measured over the inner 3, 4 and 5 deg$^2$  respectively (see dotted lines in right panel of Figure~\ref{fig-noise}). We notice that border effects are present in the very external mosaic region characterized by noise values larger than $\sim330$ $\mu$Jy (i.e. $30\times11$ 
$\mu$Jy, see last contour in Fig.~\ref{fig-noise}, left panel).

\section{The 1.4 GH\lowercase{z} Source Catalogue}
  \label{sec-extraction}

The source extraction was performed over the entire mosaic (up to rms noise values of $\sim500$ $\mu$Jy), even though the source catalogue should be considered reliable and complete only up to local noise values of 330 $\mu$Jy. 
To take into proper account both local and radial noise variations, sources were extracted from a signal-to-noise map produced by dividing the mosaic by its noise map. A preliminary list of more than 6000 sources with S/N $\geq$ 5 was derived using the MIRIAD task IMSAD. 

All the source candidates were  visually inspected. The goodness of Gaussian fit parameters was checked following \citet[][see their Sect. 2]{Prandoni00b}. Typical fitting problems arise whenever:
   \begin{itemize}
      \item Sources are fitted by IMSAD with a single Gaussian but are better described by two or more Gaussian;
      \item Sources are extended and are not well described by a Gaussian fit.
   \end{itemize}

In the first case, sources were re-fitted using multiple Gaussian components. The number of successfully split sources is 74 in total (62 in two components, 10 in three components and 2 in four components). In the second case (134 non-Gaussian sources or source components), integrated flux densities were measured by summing all the pixels above a reference 3$\sigma$ threshold, using the MIRIAD task CGCURS, which also gives the position and flux density of the source peak. Non-Gaussian sources are flagged as '{\it E}' in the catalogue. In a few additional cases Gaussian fits were able to provide good values for positions and peak flux densities, but did fail in determining the integrated flux densities. This happens typically at low signal-to-noise values. Gaussian sources with a poor determination of the integrated flux are flagged in the catalogue as '{\it G*}'. We also noticed that in a few cases our procedure (IMSAD+SExtractor) failed to detect very extended low surface brightness sources. This is due to the fact that the source itself can affect the local noise computation, producing a too high detection threshold (5$\sigma_{local}$). The few missing large low-surface brightness sources were easily recognized by eye, and added to the catalogue.

Once the final source list was produced, we computed the local noise, $\sigma_{local}$, around each source (measured in $50\times 50$ pixel regions centered at each source position in the noise map) and used it to transform the peak and integrated flux densities from S/N units to milliJy units. 

 \begin{figure}
   \centering
   \resizebox{9cm}{!}{\includegraphics[trim=0 200 0 150]{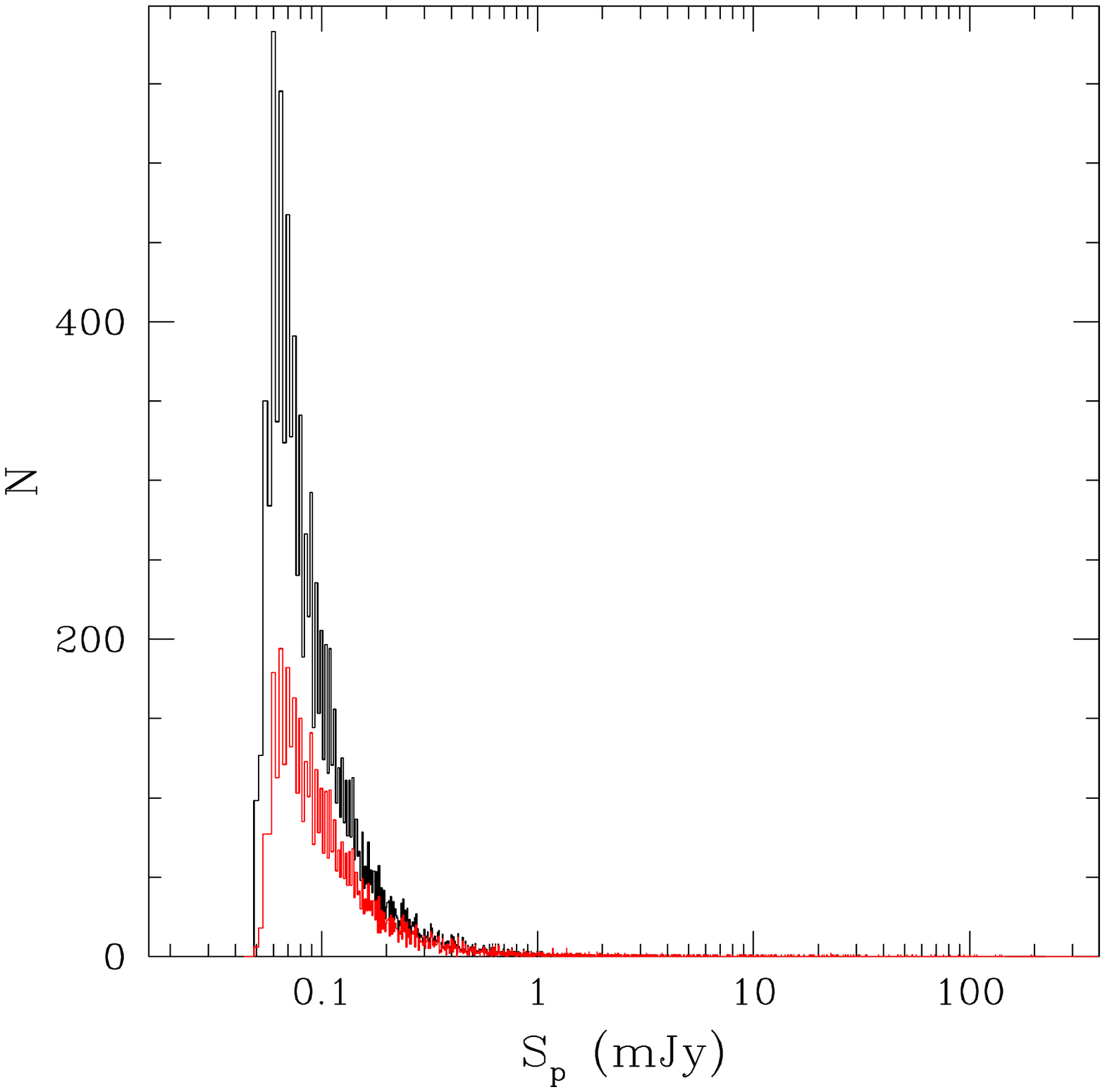}}
      \caption{Peak flux density distribution of the radio sources (or source components) before (red) and after (black) taking into proper account the noise variations along the mosaic. In the latter case the source number is weighted for the reciprocal of the visibility function shown in Figure~\ref{fig-noise} (right panel), where we assume $S_{peak} >5\sigma_{local}$ .}
         \label{fig-fluxhist}
   \end{figure}
  
  \begin{figure}
   \centering
   \resizebox{9cm}{!}{\includegraphics[trim=0 0 0 280]{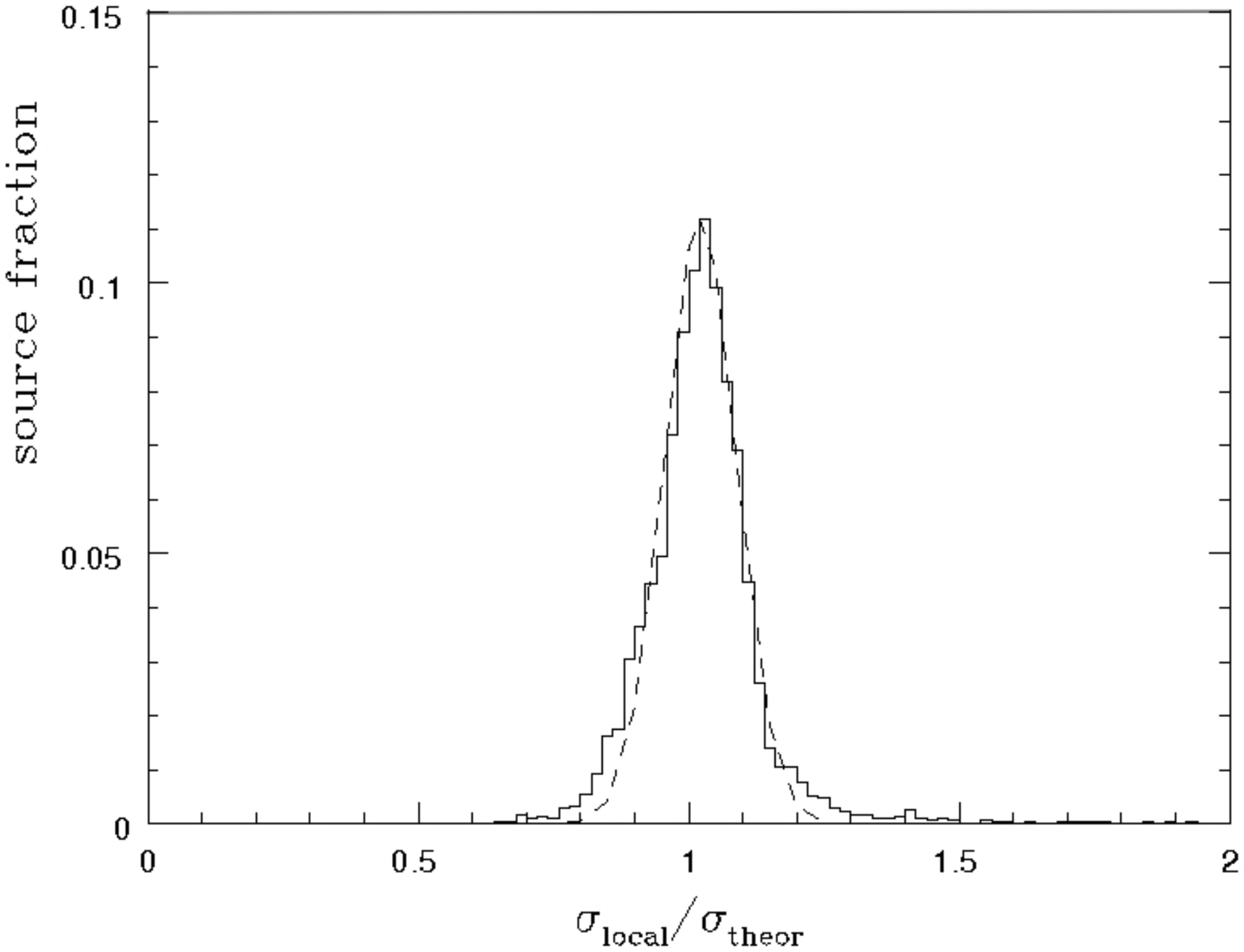}}
   \resizebox{8.5cm}{!}{\includegraphics[trim=0 0 0 250]{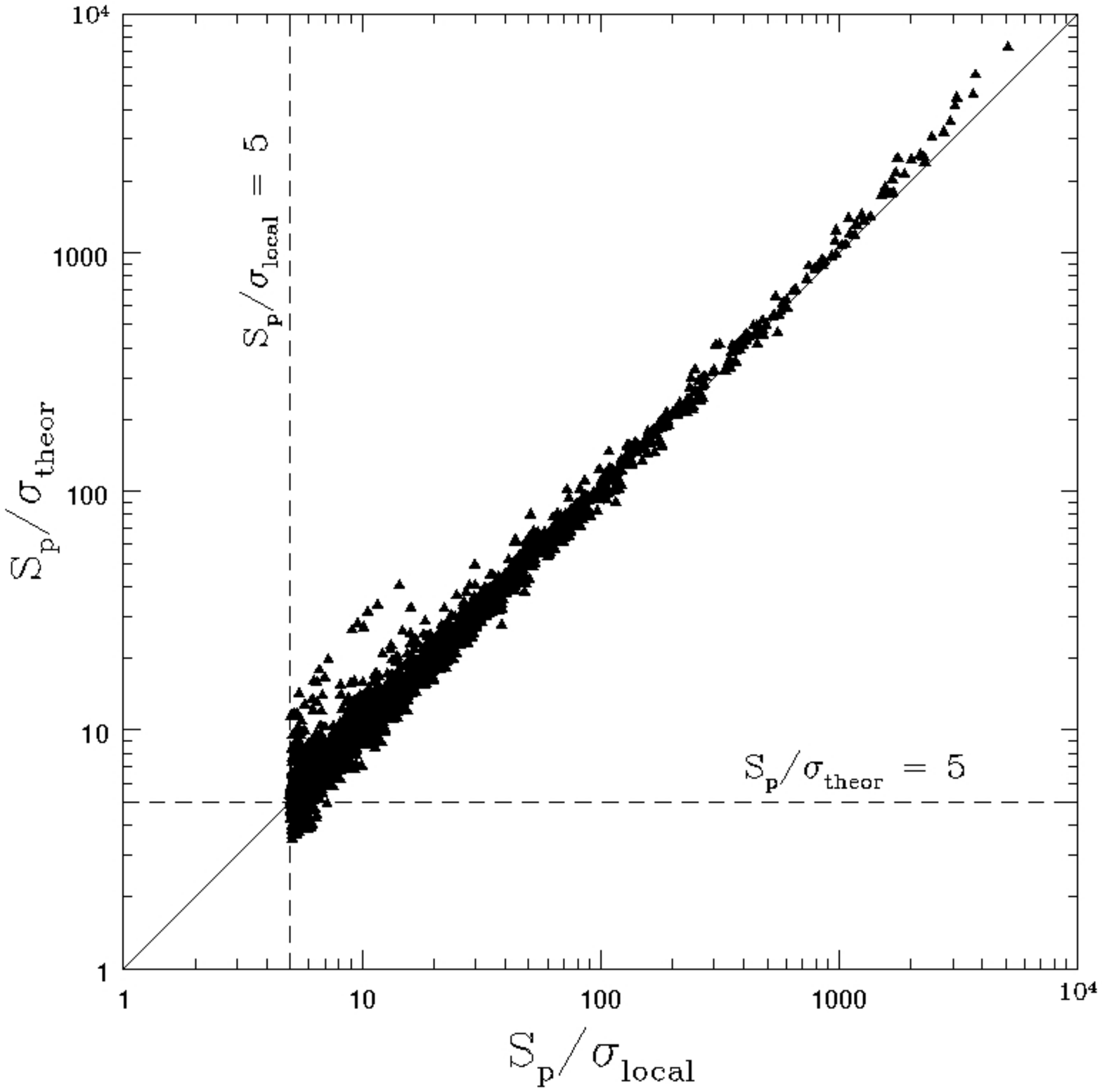}}
   \vspace{-2cm}
   \caption{\emph{Top}: Local to expected noise ratio distribution as measured in a $50\times50$ pixel box around each source. The distribution is well fitted by a Gaussian with 
   FWHM=0.137 and peak position equal to 1.02 (dashed line). \emph{Bottom}: Signal-to-noise ratios as measured using either $\sigma_{local}$ or $\sigma_{theor}$. Vertical and horizontal dashed lines indicate the  5$\sigma$ cut-off for the two signal-to-noise measurements respectively.}
              \label{fig-noiseratio}
    \end{figure}

After accounting for the splitting in multiple Gaussian components the catalog lists 6194 sources (or sources components). The peak flux distribution of our sources is shown in Figure~\ref{fig-fluxhist} before (red histogram) and after (black histogram) taking into proper account the noise variations along the mosaic. Once corrected for the source visibility function (see Fig.~\ref{fig-noise}), the peak flux distribution gets narrower, showing a steeper increase going to lower flux densities. However some incompleteness can still be seen in the lowest flux density bins. This incompleteness is the expected effect of the noise at the source extraction threshold. Due to its Gaussian distribution, whenever a source falls on a noise dip, either the source flux is underestimated or the source goes undetected. This produces incompleteness in the faintest bins. As a consequence, the measured fluxes of detected sources are biased toward higher values in the incomplete bins, because only sources that fall on noise peaks have been detected and measured. As demonstrated through Monte Carlo simulations in \citet{Prandoni00b}, incompleteness can be as high as 50\% at the 5$\sigma$ threshold, reducing down to 15\% at 6$\sigma$, and to 2\% at 7$\sigma$.  Correspondingly source fluxes are boosted by a factor of 18\% at 5$\sigma$, of 10\% at 6$\sigma$ and of 6\% at 7$\sigma$. However such incompleteness effects can be counterbalanced (at least partially -- the extent actually depends on the shape of the source counts) by the fact that sources below the detection threshold can be pushed above it when they sit on a noise peak. 

\subsection{Noise Analysis}\label{sec-artefacts}

To better investigate the local noise  ($\sigma_{local}$) distribution, we compared it with the expected noise ($\sigma_{theor}$), defined as the average noise value measured within a $50\times50$ pixel box centered at the same position in the so-called \emph{sensitivity map}, which is a map of the expected noise, based on the integration time spent on each observed field, and on the complex primary beam response obtained when linearly combining all the fields in the final mosaic. 
As shown in Figure~\ref{fig-noiseratio} (top panel), the local noise does not generally shows significant systematic departures from the expected rms value: the distribution can be described fairly well by a Gaussian with FWHM=0.137 and a peak position equal to 1.02 (dashed line).
Also compared are the signal-to-noise ratios defined, for each source, using either $\sigma_{local}$ or  $\sigma_{theor}$ (Figure~\ref{fig-noiseratio}, bottom panel). The two measured signal-to-noise ratios mostly agree with each other, although a number of significant departures are evident for the faintest and brightest sources. This is due to the presence of some residual areas where the noise is not random due to systematic effects (typically phase errors around bright sources). 

\begin{figure}
   \centering
   \resizebox{8cm}{!}{\includegraphics{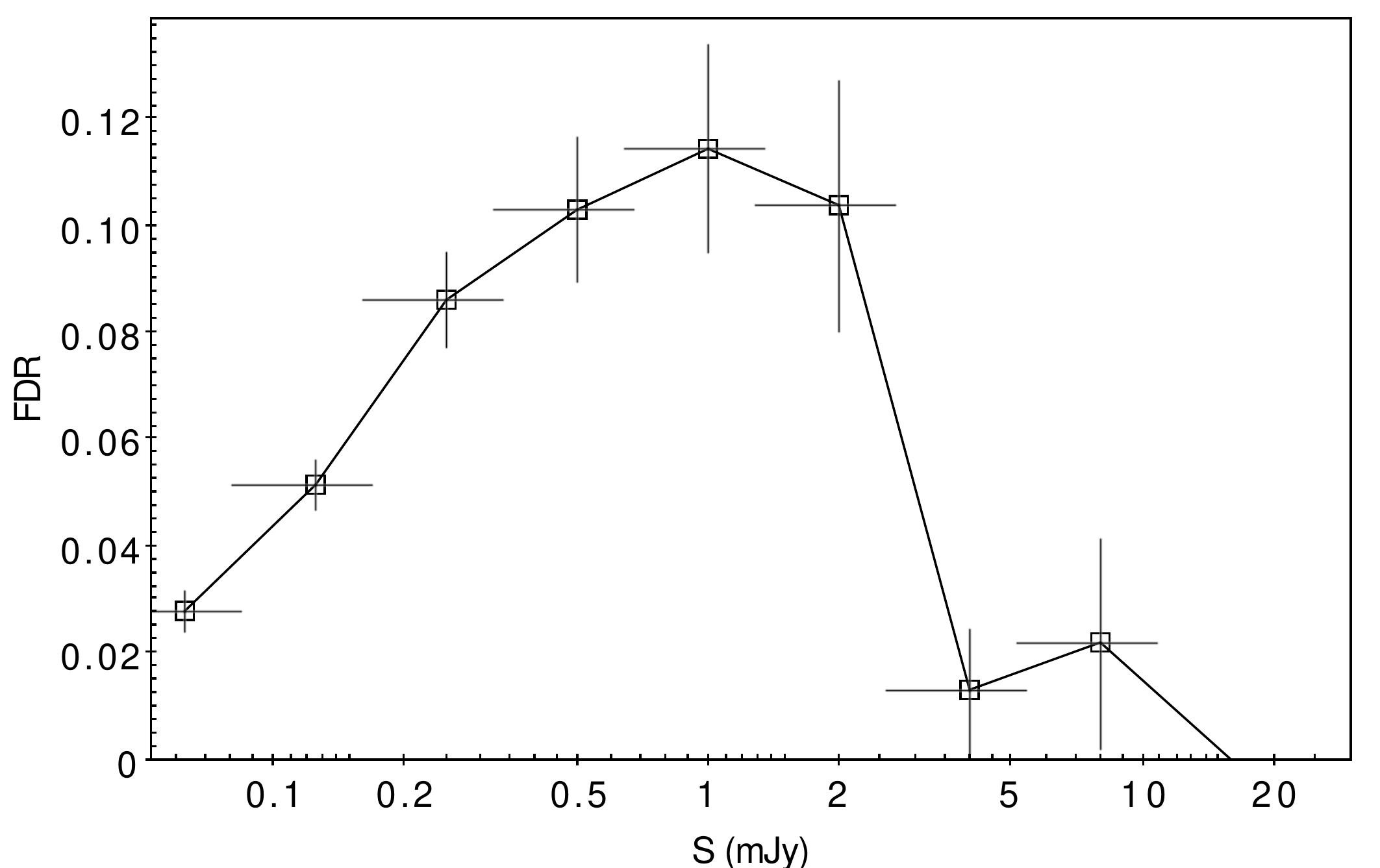}}
   \caption{False detection rate as a function of integrated flux density. }
              \label{fig-fdr}
    \end{figure}

To quantify the effect of non-Gaussian noise on our source catalogue, we quantified the number of possible spurious detections  in the following way. By assuming that negative and positive noise spikes have a similar distribution, we ran IMSAD on the negative mosaic map (i.e. the map multiplied by -1), with the same input parameters used to extract the source catalogue. We found 356 components above the 5$\sigma$ threshold, within the completeness area of the catalogue (local noise $<330$ $\mu$Jy), corresponding to a fraction of 5.8\%. The false detection rate (FDR; i.e. the ratio between the number of spurious components and the number of components in the catalogue) as a function of total flux is shown in Fig.~\ref{fig-fdr}. The FDR peaks around  $\sim 0.5 - 2$ mJy, where we can expect a contamination from artefacts $\gtrsim10\%$.
Sources which from visual inspection appear to be likely noise peak are flagged as '{\it n}' in the catalogue.  

\subsection{Bandwidth smearing}

Bandwidth smearing, the radio analog of optical chromatic aberration, is a well-known effect caused by the finite width of the receiver channels. It reduces the peak flux density of a source while correspondingly increasing the apparent source size in the radial direction such that the total integrated flux density is conserved. The amount of smearing is proportional to the distance from the phase center and the channel width (or passband) of the data. Assuming a Gaussian beam and passband (see \citealt{Condon98}), we find that in our particular case the expected peak flux density attenuation at the maximum distance from the phase center  (50 arcmin; see Sect.~\ref{sec-observations}) is $S_{peak}$/$S^{0}_{peak}$=$0.999$, where $S^{0}_{peak}$ represents the un-smeared  source peak flux density. It is therefore clear that bandwidth smearing is not an issue for our source catalogue.


\subsection{Deconvolution}\label{sec-deconv}

\begin{figure}
   \centering
   \resizebox{9cm}{!}{\includegraphics[trim=0 0 0 130]{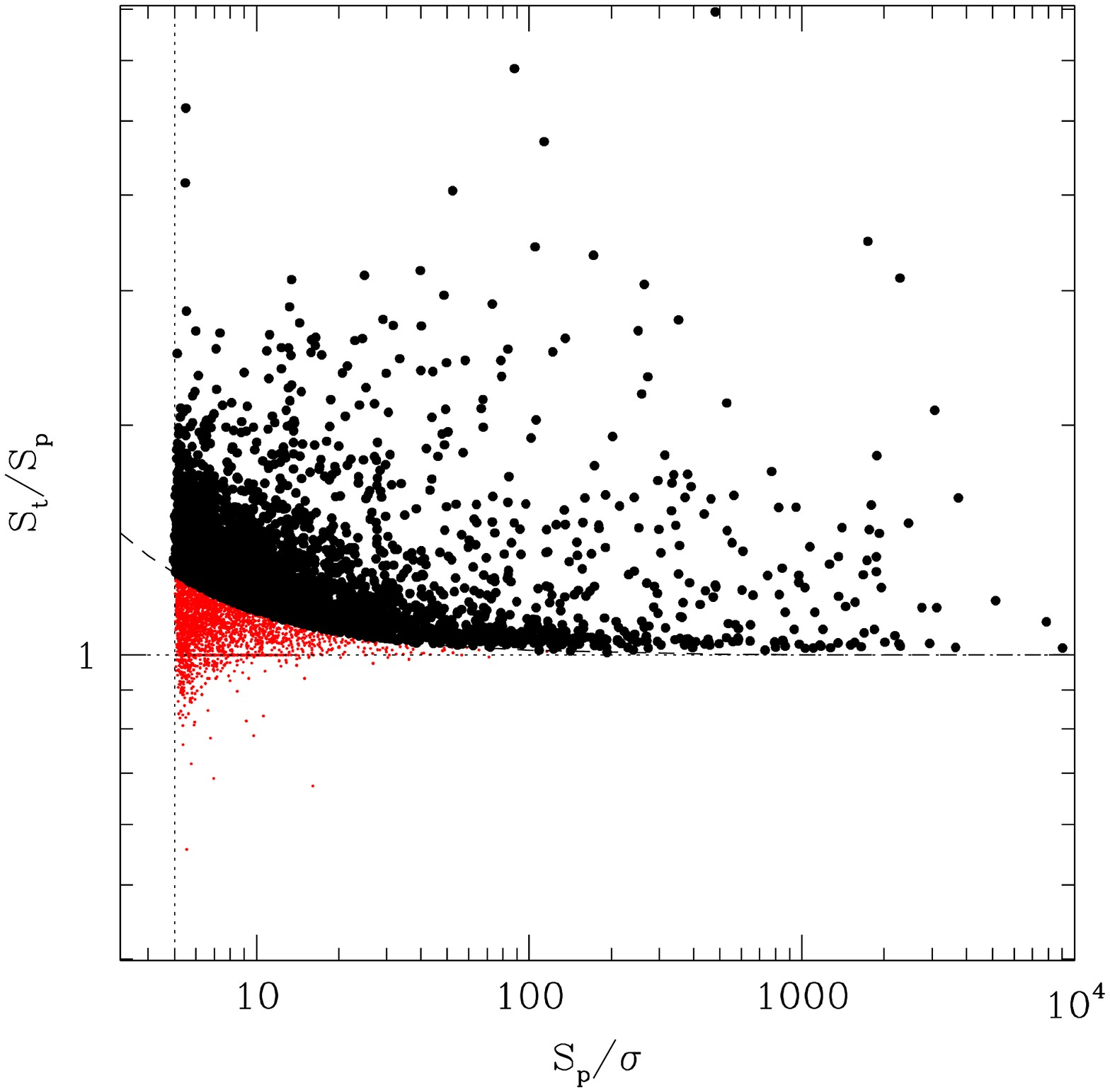}}
   \vspace{-2cm}
      \caption{Integrated  to peak flux ratio as a function of the source signal-to-noise. Lines indicate the S/N=5 cut-off adopted in the source catalogue extraction (dotted), the locus $S_t$/$S_p$=1 (solid), and  the envelop function defined by Eq.~(\ref{eq-ratio}) (dashed). Red dots correspond to unresolved sources; black filled circles correspond to resolved or partially resolved sources (i.e. those sources that are resolved only along the major axis).}
         \label{fig-sratio}
   \end{figure}

   The ratio of the integrated flux to the peak flux is a direct measure of the extent of a radio source:
    \begin{equation}
      S_{tot}$/$S_{peak} $=$ \theta_{maj}\theta_{min}$/$b_{maj}b_{min}
      \label{eq-sratio}
   \end{equation}
   where  $\theta_{maj}$  and $\theta_{min}$  are the source FWHM axes and $b_{maj}$ and $b_{min}$  are the synthesized beam FWHM axes. The flux ratio can therefore be used to discriminate between extended (larger than the beam) and point-like sources. In Figure~\ref{fig-sratio} we have plotted the flux ratio $S_{tot}$/$S_{peak}$ as a function of the signal-to-noise for all the sources (or source components) in our catalogue. 

   The flux density ratio has a skewed distribution, with a tail towards high flux ratios due to extended sources. To establish a criterion for classifying extended sources, errors in the flux measurement have to be taken into account, since such errors can introduce an intrinsic spread even in case of points sources. We have determined the 1$\sigma$ error fluctuation of the  ratio $S_{tot}$/$S_{peak}$ as a function of the signal-to-noise ratio using the \citet{Condon97} equations of error propagation derived for two dimensional elliptical Gaussian fits of point sources in presence of Gaussian noise (see Eqs.~\ref{eq-errflux}, \ref{eq-errmaj}, \ref{eq-errmin} in Sect.~\ref{sec-errors}). 
   We find an envelope function that can be characterized by the equation:
    \begin{equation}
      S_{tot}$/$S_{peak} $=$ 1 $+$ 1.4$($\frac{S_{peak}}{\sigma_{local}}$)$^{-1}
  \label{eq-ratio}
   \end{equation}
(see dashed line in Fig.~\ref{fig-sratio}).
   
We have then considered as truly resolved only those sources laying above such envelope. From this analysis we found that 2548 sources (or source components) in the catalogue are  unresolved (red dots in Fig.~\ref{fig-sratio}). Another 599 are resolved only in the major axis direction. In total we have 3047 fully resolved sources ($\sim 50\%$ of the sample). The deconvolved angular sizes of unresolved sources are set to zero in the catalogue.  For a size distribution of the sources we refer to Figure~\ref{fig-sizeflux}.
It is worth noting that the fraction of unresolved/resolved sources in a radio catalogue very much depends on the criteria adopted to make this distinction. Assuming a more conservative envelope function that accounts for 2$\sigma$ $S_{tot}$/$S_{peak}$ fluctuations,  the fraction of resolved sources would get down to 27\%. 

\subsection{Multiple-Component Sources}
   
\begin{figure}
   \centering
   \resizebox{9.5cm}{!}{\includegraphics[trim=0 0 0 110]{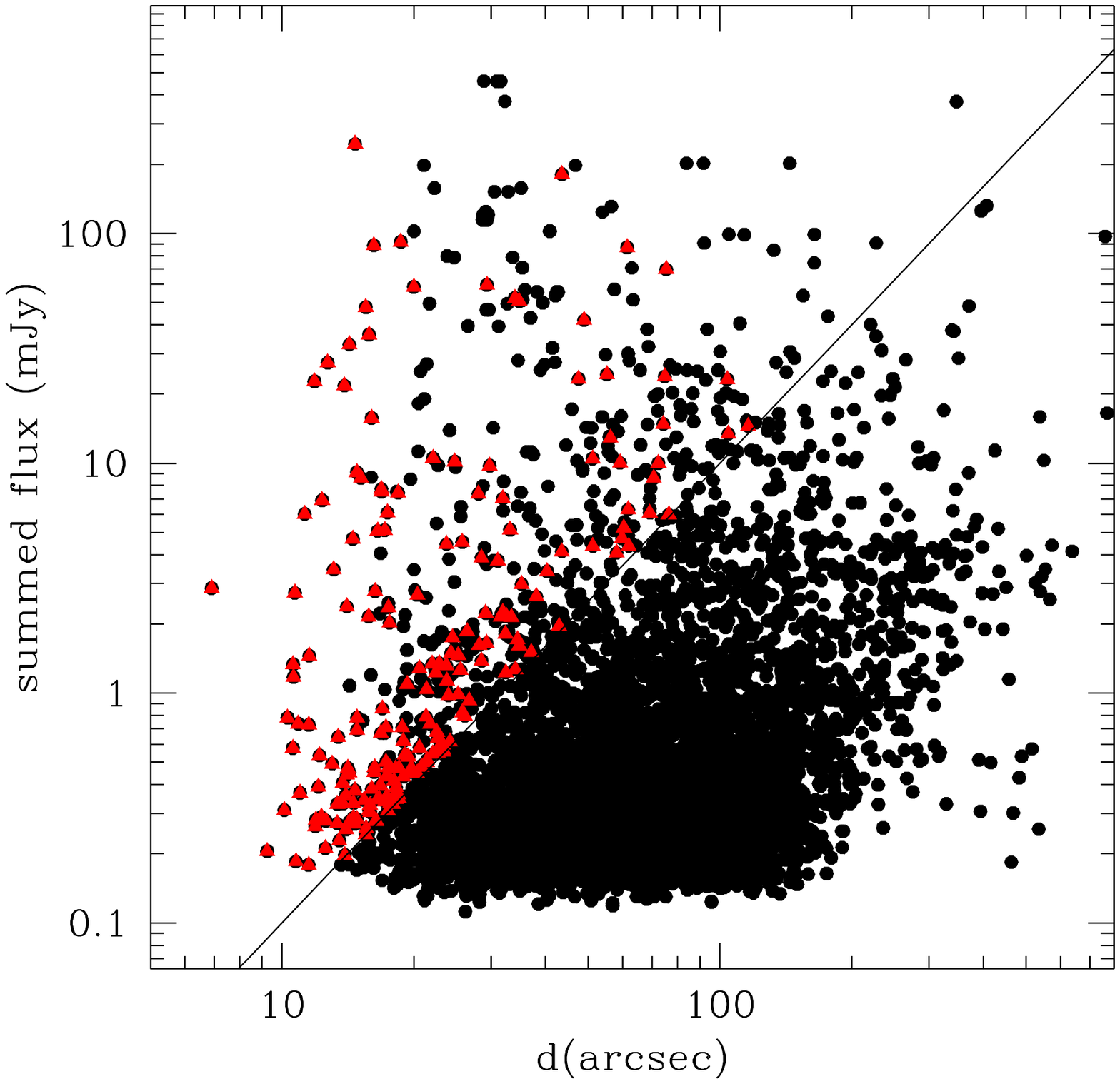}}
  \vspace{-2.9cm}
      \caption{Sum of the flux densities of nearest neighbor pairs plotted against their separation. Source pairs that lie above the solid line and have flux densities that differ by less than a factor of 4 are considered as double source candidates (see red filled triangles). }
         \label{fig-multiple}
   \end{figure}

Radio sources associated with radio galaxies can be made up of a nucleus with hot spots along, or at the end of, one or two jets. The individual components of a single source are often cataloged separately by Gaussian fitting routines, so a method must be devised to identify multiple components as belonging to a single source.When jets are detected it is relatively easy to recognize the components belonging to the same source, and indeed through visual inspection we were able to recognize such cases (see discussion above). More difficult is the case when only lobes are detected (double-component sources), and no sign of connecting jets is present.
To recognize such double--component sources we applied the statistical technique of \citet{Magliocchetti98}, later modified by \citet{Huynh05}, where the sum of the fluxes of each nearest neighbor pair ($S_{sum}$) versus their separation (d) is analyzed (see Figure~\ref{fig-multiple}). The high density of points to the lower right of the  $S_{sum}$ -- $d$ plane is to be ascribed to the general population of single-component sources. 
   Following \citet{Huynh05} a maximum allowed separation for double-component sources is then applied as a function of the summed flux density, as follows:
\begin{equation}
      d_{max}$=$100 $($S_{sum}$/$10$)$^{0.5}
      \label{eq-multi}
\end{equation}
where $S_{sum}$ is given in mJy and $d_{max}$ in arcsec. This maximum separation is shown as a solid line in Figure~\ref{fig-multiple}.
   This procedure proves to be more successful than distance-only based criteria for very deep surveys like ours, where random pairs can be found even at very small separations. In fact a maximum separation varying with summed flux allows to consider \emph{faint} pairs, which are likely to be found by chance, as single sources even when at very small separations; at the same time it allows to include among real pairs \emph{bright} sources at large separation.

  Since flux densities of components of real double sources are typically similar, a second constraint was then applied to restrict the matched pairs to real physically associated sources. Following \citet{Huynh05}, we consider pairs as really physically associated only if their flux densities differ by a factor of less than 4. Sources that meet this further requirement are shown in Figure~\ref{fig-multiple} as large filled triangles. 
From this analysis we could identify 155 additional double source candidates. All such sources were again visually inspected, and 46 were discarded as they are clearly random associations (mainly based on  component morphology and pair alignment considerations). The  remaining 109 pairs are included as double sources and flagged '{\it M}' in this release of the catalogue. In the future these sources will be further investigated (and possibly confirmed as multiple) through inspection of the deep optical/infrared catalogues and images covering the LH region. This will allow us to identify the host galaxy possibly associated to (multiple) radio sources.  

\subsection{Catalog Format}
The final catalog lists 5997 sources, including 183 multiple-component sources. Most (90\%) are sub-mJy sources. The  full radio catalogue is available in electronic form.   
A sample  is shown in Table~\ref{tab-Catalog}. 

For multiple sources we list all the components (labeled 'A', 'B', etc.) preceded by a line (flagged '{\it M}') giving the position of the radio centroid, the total flux density and the overall angular size of the source. Source positions have been defined as the flux-weighted average position of the components (source centroid). For sources with more than two components the centroid position has been replaced with the core position whenever the core is clearly recognizable. Total source flux densities are computed by summing all the component integrated fluxes. Multiple source angular sizes are defined as largest angular sizes ({\it las}), i.e. the maximum distance between the source components. 

 \begin{table*}
      \caption{The radio catalogue: sample (the full version is available in electronic form).The format is the following:
 \emph{Column (1)}: Source IAU name. The components of multiple sources are labeled 'A', 'B', etc.
 \emph{Column (2) and (3)}: Source position: Right Ascension and Declination (J2000).
 \emph{Column (4)}: Source 1.4 GHz peak flux density, in mJy.
\emph{Column (5)}: Source 1.4 GHz integrated flux density, in mJy.
\emph{Column (6) and (7)}: Fitted major and minor axes (FWHM) of the source in arcsec.
 \emph{Column (8)}: Fitted position angle (P.A., measured N through E) for the major axis in degrees.
 \emph{Column (9) and (10)}:  Deconvolved major and minor axes (FWHM) of the source in arcsec. Zero values refer to unresolved (or partially unresolved) sources.
 \emph{Column (11)}: Deconvolved position angle (P.A., measured N through E) for the major axis in degrees. Zero values refer to unresolved sources.
\emph{Column (12)}: Source local noise ($\sigma_{local}$) in mJy.
\emph{Column (13)}: Fitting Flag. Flag indicating the fitting procedure and parameterization adopted for the source or source component. {\it G} refers to Gaussian fit.  {\it E} refers to non-Gaussian sources.  {\it M} refers to global parameters of multiple sources (see text for more details).
 \emph{Column (14)}:  Morphology flag. An additional flag is given in some specific cases: 1) {\it *} when the Gaussian fit is poor (see Sect.~\ref{sec-extraction} for more details); 2) {\it c} when the source is well fitted by a Gaussian but shows signs of a more {\it complex} morphology; 3) {\it m} when the source is catalogued as a single source, but shows signs of {\it multiple} components; 4) {\it n} when a source appears to be spurious ({\it noise} artifact). 
}
         \label{tab-Catalog}
      \scriptsize  
      \begin{tabular}{lllrrrrrrrrrlc}
      \hline
      \multicolumn{1}{c}{IAU Name} & \multicolumn{1}{c}{RA} & \multicolumn{1}{c}{Dec} & \multicolumn{1}{c}{$S_{peak}$} &  \multicolumn{1}{c}{$S_{tot}$} &  \multicolumn{1}{c}{$\theta_{maj}$} &  \multicolumn{1}{c}{$\theta_{min}$} & \multicolumn{1}{c}{P.A.}&  \multicolumn{1}{c}{$d\theta_{maj}$} &  \multicolumn{1}{c}{$d\theta_{min}$} &  \multicolumn{1}{c}{dP.A.} &  \multicolumn{1}{c}{$\sigma_{local}$} &    &\\
      \multicolumn{1}{c}{} & \multicolumn{1}{c}{J2000} & \multicolumn{1}{c}{J2000} &  \multicolumn{1}{c}{(mJy)} &  \multicolumn{1}{c}{(mJy)} &  \multicolumn{1}{c}{(arcsec)} &  \multicolumn{1}{c}{(arcsec)} &  \multicolumn{1}{c}{degr} &  \multicolumn{1}{c}{(arcsec)} &  \multicolumn{1}{c}{(arcsec)} &  \multicolumn{1}{c}{degr} &  \multicolumn{1}{c}{(mJy)} &  &\\
      \hline 
 & & & & & & & & & & & & & \\
 LHW J104515+580634 & 10 45 15.31 & +58 06 34.6 &  1.099 &  1.323 & 12.10 &  9.84 & 27.4 &  6.45 &  0.00 & 52.9 & 0.0301 & G & \\
 LHW J104515+575711 & 10 45 15.92 & +57 57 11.4 &  0.321 &  0.574 & 18.10 &  9.78 & -2.5 & 14.38 &  3.82 & -3.0 & 0.0262 & G & c \\
 LHW J104517+580737 & 10 45 17.82 & +58 07 37.1 &  0.281 &  0.301 & 11.50 &  9.20 & 24.2 &  0.00 &  0.00 &  0.0 & 0.0295 & G & \\
 LHW J104518+571626 & 10 45 18.41 & +57 16 26.7 &  0.233 &  0.267 & 12.80 &  8.82 & 16.7 &  0.00 &  0.00 &  0.0 & 0.0293 & G & \\
 LHW J104518+571012 & 10 45 18.90 & +57 10 12.9 &  0.184 &  0.236 & 12.40 & 10.22 & 12.6 &  6.23 &  4.17 & 38.8 & 0.0330 & G & \\
 & & & & & & & & & & & & & \\
 LHW J104519+581742 & 10 45 19.19 & +58 17 42.5 &  0.151 &  0.148 & 10.70 &  9.03 & -2.8 &  0.00 &  0.00 &  0.0 & 0.0288 & G & \\
 LHW J104519+581044 & 10 45 19.91 & +58 10 44.6 &  0.157 &  0.210 & 13.80 &  9.58 & -15.2 &  8.59 &  2.54 & -24.0 & 0.0298 & G & \\
 LHW J104519+571545 & 10 45 19.97 & +57 15 45.5 &  0.245 &  0.282 & 11.80 &  9.60 &  0.4 &  0.00 &  0.00 &  0.0 & 0.0290 & G & \\
 LHW J104520+581224 & 10 45 20.54 & +58 12 24.1 &  0.345 &  0.417 & 11.70 & 10.17 &  8.3 &  5.04 &  3.60 & 64.8 & 0.0291 & G & \\
 LHW J104520+583714 & 10 45 20.60 & +58 37 14.8 &  0.219 &  0.232 & 11.80 &  8.83 &  1.8 &  0.00 &  0.00 &  0.0 & 0.0284 & G & \\
  & & & & & & & & & & & & & \\
 LHW J104521+575553 & 10 45 21.87 & +57 55 53.4 &  0.154 &  0.221 & 13.70 & 10.33 & -24.5 &  8.80 &  3.87 & -38.9 & 0.0264 & G & \\
 LHW J104521+575027 & 10 45 21.90 & +57 50 27.8 &  0.200 &  0.202 & 10.70 &  9.34 & 14.5 &  0.00 &  0.00 &  0.0 & 0.0262 & G & \\
 LHW J104522+574827 & 10 45 22.31 & +57 48 27.3 & 12.450 & 14.829 & 39.84 & 23.96 & 48.1 & 38.58 & 21.69 & 49.2 & 0.0263 & E & m \\
 LHW J104522+572824 & 10 45 22.44 & +57 28 24.6 &  0.158 &  0.195 & 12.80 &  9.50 & -13.1 &  0.00 &  0.00 &  0.0 & 0.0281 & G & \\
 LHW J104522+582203 & 10 45 22.54 & +58 22 03.9 &  0.172 &  0.178 & 10.50 &  9.70 &  2.4 &  0.00 &  0.00 &  0.0 & 0.0257 & G & \\
  & & & & & & & & & & & & & \\
 LHW J104522+590742 & 10 45 22.59 & +59 07 42.7 &  2.585 &  2.421 & 10.40 &  8.85 & -13.4 &  0.00 &  0.00 &  0.0 & 0.3610 & G & \\
 LHW J104522+571727 & 10 45 22.63 & +57 17 27.9 &  0.164 &  0.177 & 12.50 &  8.53 & 22.7 &  0.00 &  0.00 &  0.0 & 0.0282 & G & \\
 LHW J104523+573057 & 10 45 23.20 & +57 30 57.9 &  0.347 &  0.458 & 13.60 &  9.56 & -7.2 &  8.06 &  3.05  & -12.3 & 0.0282 & G &  \\
 LHW J104523+580913 & 10 45 23.48 & +58 09 13.1 &  0.431 &  0.634 & 12.40 & 11.69 & -18.8 &  7.64 &  5.48 & -79.2 & 0.0288 & G & \\
 LHW J104524+582957 & 10 45 24.00 & +58 29 57.5 &  0.895 &  0.937 & 10.90 &  9.46 & -12.7 &  3.52 &  0.00 & -68.5 & 0.0252 & G & \\
  & & & & & & & & & & & & & \\
 LHW J104524+582610 & 10 45 24.31 & +58 26 10.7 &  0.183 &  0.243 & 14.10 &  9.33 & -1.6 &  8.82 &  2.45 & -2.5 & 0.0240 & G & \\
 LHW J104524+575926 & 10 45 24.34 & +57 59 26.8 &  0.206 &  0.234 & 12.20 &  9.22 & -23.6 &  0.00 &  0.00 &  0.0 & 0.0260 & G & \\
 LHW J104524+573831 & 10 45 24.89 & +57 38 31.0 &  0.169 &  0.156 & 11.20 &  8.07 & 14.6 &  0.00 &  0.00 &  0.0 & 0.0313 & G & n \\
 LHW J104524+570933 & 10 45 24.92 & +57 09 33.1 &  0.494 &  0.511 & 11.30 &  9.01 &  0.5 &  0.00 &  0.00 &  0.0 & 0.0309 & G & \\
 LHW J104525+573808 & 10 45 25.69 & +57 38 08.6 &  0.329 &  0.441 & 13.00 & 10.19 & -34.7 &  8.19 & 1.94 & -52.8 & 0.0315 &G & n \\
 & & & & & & & & & & & & & \\ 
 LHW J104526+583531 & 10 45 26.77 & +58 35 31.1 &  0.582 &  0.788 & 32.70 &       &       &       &      &       & 0.0242 & M &   \\
 LHW J104526+583531A & 10 45 26.76 & +58 35 26.6 &  0.582 &  0.620 & 11.10 &  9.43 & -3.1 &  2.93 &  1.26 & -74.0 & 0.0242 & G & \\
 LHW J104526+583531B & 10 45 26.78 & +58 35 47.9 &  0.137 &  0.168 & 11.70 & 10.38 & -4.2 &  0.00 &  0.00 &  0.0 & 0.0243 & G & \\
 LHW J104527+572436  & 10 45 27.61 & +57 24 36.9 &  0.307 &  0.390 & 13.40 &  9.33 & -16.9 &  8.00 &  0.81 & -27.2 & 0.0247 & G & \\
 LHW J104527+564137  & 10 45 27.84 & +56 41 37.2 &  2.200 &  2.167 & 10.90 &  8.89 &  1.6 &  0.00 &  0.00 &  0.0 & 0.4140 & G & \\
 & & & & & & & & & & & & & \\ 
 LHW J104527+565910 & 10 45 27.93 & +56 59 10.0  & 0.301  & 0.268  & 10.40 &  8.43 & 31.6 &  0.00 &  0.00 &  0.0 & 0.0506 & G & \\
 LHW J104528+572928 & 10 45 28.36 & +57 29 28.2  & 10.521 & 22.671 & 40.63 &       &      &       &       &      & 0.0272 & M & \\
 LHW J104528+572928A & 10 45 27.74 & +57 29 28.2 & 10.521 & 13.078 & 11.29 & 10.90 & 26.8 &  6.32 &  2.09 & 84.4 & 0.0272 & G & \\
 LHW J104528+572928B & 10 45 29.21 & +57 29 28.2 &  7.968 &  9.593 & 11.31 & 10.54 & -23.8 &  5.84 &  1.70 & -78.3 & 0.0270 & G & \\
 LHW J104528+575347 & 10 45 28.45 & +57 53 47.5  & 0.143  & 0.192  & 11.70 & 11.36 & 32.9 &  7.12 &  3.64 & 84.5 & 0.0259 & G & \\
  & & & & & & & & & & & & & \\
 LHW J104529+581749 & 10 45 29.28 & +58 17 49.2  & 0.129  & 0.159  & 13.40 &  9.06 & 12.8 &  0.00 &  0.00 &  0.0 & 0.0254 & G & \\
 LHW J104529+573817 & 10 45 29.97 & +57 38 17.0  & 66.508 & 70.508 & 11.40 &  9.21 &  7.2 &  3.51 &  0.69 & 35.8 & 0.0303 & G & \\
 LHW J104530+581231 & 10 45 30.47 & +58 12 31.6  & 1.791  & 1.945  & 11.40 &  9.38 &  4.5 &  3.37 &  2.15 & 38.7 & 0.0260 & G & \\
 LHW J104530+571220 & 10 45 30.52 & +57 12 20.0  & 0.188  & 0.270  & 15.30 &  9.26 & -3.9 & 10.65 &  2.12 & -5.3 & 0.0270 & G & \\
 LHW J104530+583828 & 10 45 30.85 & +58 38 28.4  &  0.515 & 0.535  & 10.70  &  9.56  & 5.4   & 0.00 &  0.00 &  0.0 & 0.0251 & G & \\ 
& & & & & & & & & & & & & \\
      \hline
      \end{tabular}
   \end{table*}


\section{Errors in Source Parameters}
\label{sec-errors}

   Parameter uncertainties are the quadratic sum of two independent terms: the calibration errors, which dominate at high signal-to-noise ratios, and the internal errors, due to the presence of noise in the maps. The latter dominate at low signal-to-noise ratios. 
For an estimate of the internal errors of the source parameters we refer to Condon's master equations \citep{Condon97}, which provide error estimates for elliptical Gaussian fitting procedures. Such equations already proved to be adequate to describe the measured internal errors for other similar deep 1.4 GHz radio catalogues, obtained with the same detection and fitting algorithm (IMSAD) applied to radio mosaics (see e.g. \citealt{Prandoni00b}). Applying Condon's master equation to our radio survey, we derived the relations which describe 1$\sigma$ internal errors for flux density and source axis fitting measurements for point sources ($\vartheta_{maj} \times \vartheta_{min}$ = $11^{\prime\prime}\times 9^{\prime\prime}$, PA= 0$^{\circ}$):
\begin{equation}
      \sigma $($S_{peak}$)$$/$S_{peak} $=$ 1.00 $($\frac{S_{peak}}{\sigma}$)$^{-1}
  \label{eq-errflux}
   \end{equation}
\begin{equation}
      \sigma $($\theta_{maj}$)$$/$\theta_{maj}) $=$ 1.11$($\frac{S_{peak}}{\sigma}$)$^{-1}   
   \label{eq-errmaj}
   \end{equation}
\begin{equation}
      \sigma $($\theta_{min}$)$$/$\theta_{min}) $=$ 1.11 $($\frac{S_{peak}}{\sigma}$)$^{-1}  
     \label{eq-errmin}
   \end{equation}
As demonstrated in \citet{Prandoni00b}, the fact that a source is extended does not affect the internal accuracy of the fitting algorithm and therefore the errors quoted above apply to fitted flux densities and source sizes of extended sources as well.

 Similar equations hold for position 1$\sigma$ internal errors \citep{Condon97,Condon98}, that applied to  point sources in our radio survey, reduce to:
\begin{equation}
      \sigma$($\alpha$)$ $=$ 3.46 $($\frac{S_{peak}}{\sigma}$)$^{-1}  \;\;\;\;\;\;\;\;\;\;\;\;\;\;\;\;\;\; (\rm{arcsec})
      \label{eq-errra}
   \end{equation}
\begin{equation}
      \sigma$($\delta$)$ $=$ 5.16 $($\frac{S_{peak}}{\sigma}$)$^{-1}  \;\;\;\;\;\;\;\;\;\;\;\;\;\;\;\;\;\; (\rm{arcsec})
      \label{eq-errdec}
   \end{equation}

Calibration terms are in general estimated from comparison with external data of better accuracy than the one tested. As discussed in Sect.~\ref{sec-multiband}, the LH region has been observed at 1.4 GHz by previous smaller surveys, using the \emph{Very Large Array}. In addition the region is covered by shallower VLA all sky surveys like the NVSS \citep{Condon98} and the FIRST \citep{Becker95}. None of such surveys can in principle be considered of better accuracy than our survey. Nevertheless a source flux density/position comparison with those samples, allows us to check the consistency of our parameter measurements and calibration with those of the other existing surveys, and check for any systematic effects that we might have introduced in the image processing (especially at low signal-to-noise values).

\subsection{Comparison with external data}

Our source catalogue was cross-correlated with the shallow NVSS and FIRST all-sky catalogues (limiting fluxes $\sim$2.5 and $\sim$1 mJy respectively) and with three other deeper overlapping catalogues (as described in Sect.~\ref{sec-multiband}):  \citet[][VLA C-configuration]{deruiter97}; \citet[][VLA mostly B-configuration]{Ibar09}; and \citet[][VLA A+B-configurations]{Biggs06}. 

\begin{figure*}
\vspace{-2cm}
\resizebox{12cm}{!}{\includegraphics[width=12cm]{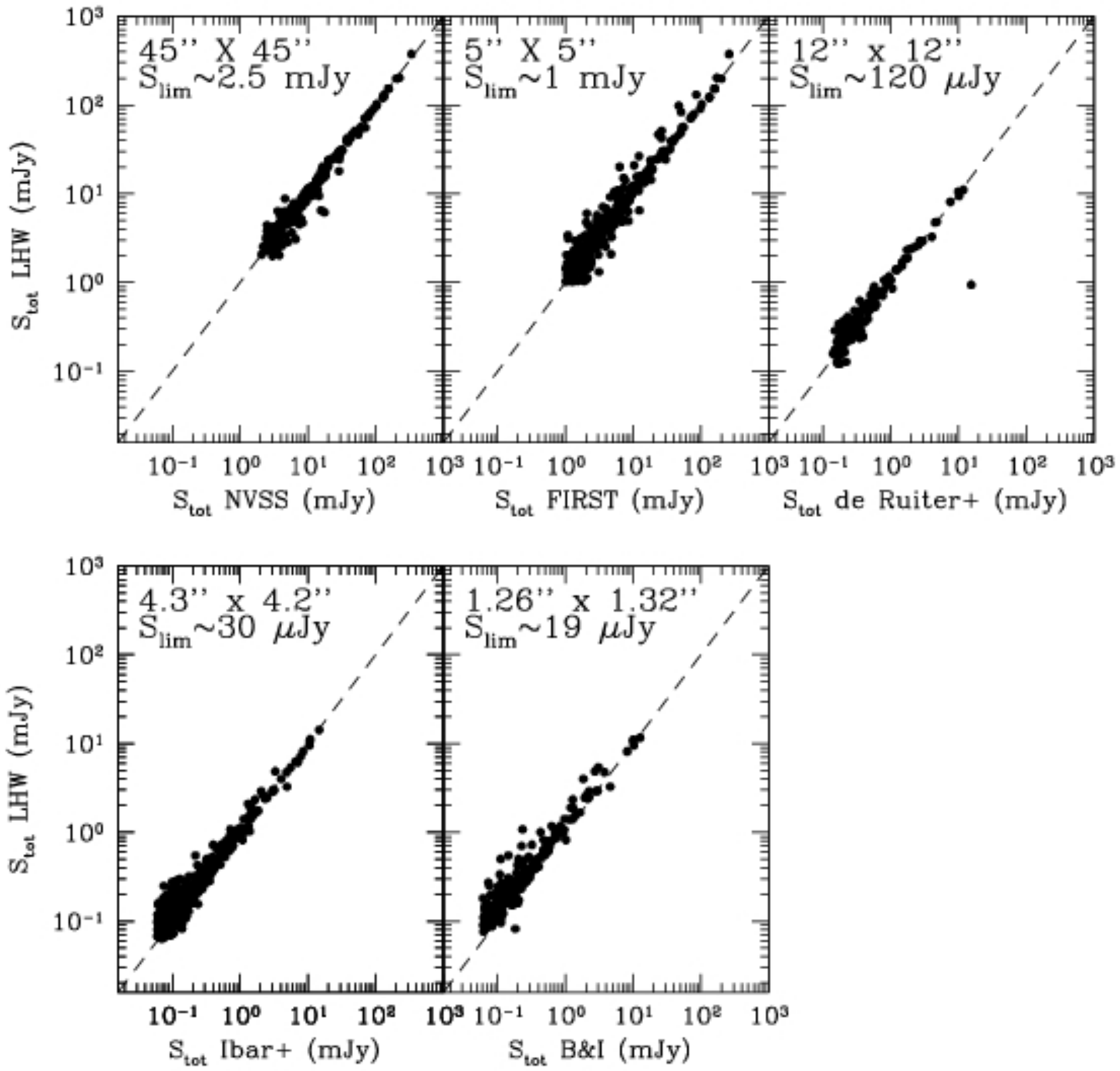}}
\hfill
\begin{minipage}[t]{55mm}
\vspace{-5cm}
\caption{Flux comparison between our sample and other existing overlapping catalogues. The limiting flux and the spatial resolution of  comparison samples are reported in  each panel. See text for more details.}
         \label{fig-confflux}
\end{minipage}
\end{figure*}

\begin{figure*}
\vspace{-2cm}
\resizebox{12cm}{!}{\includegraphics[width=12cm]{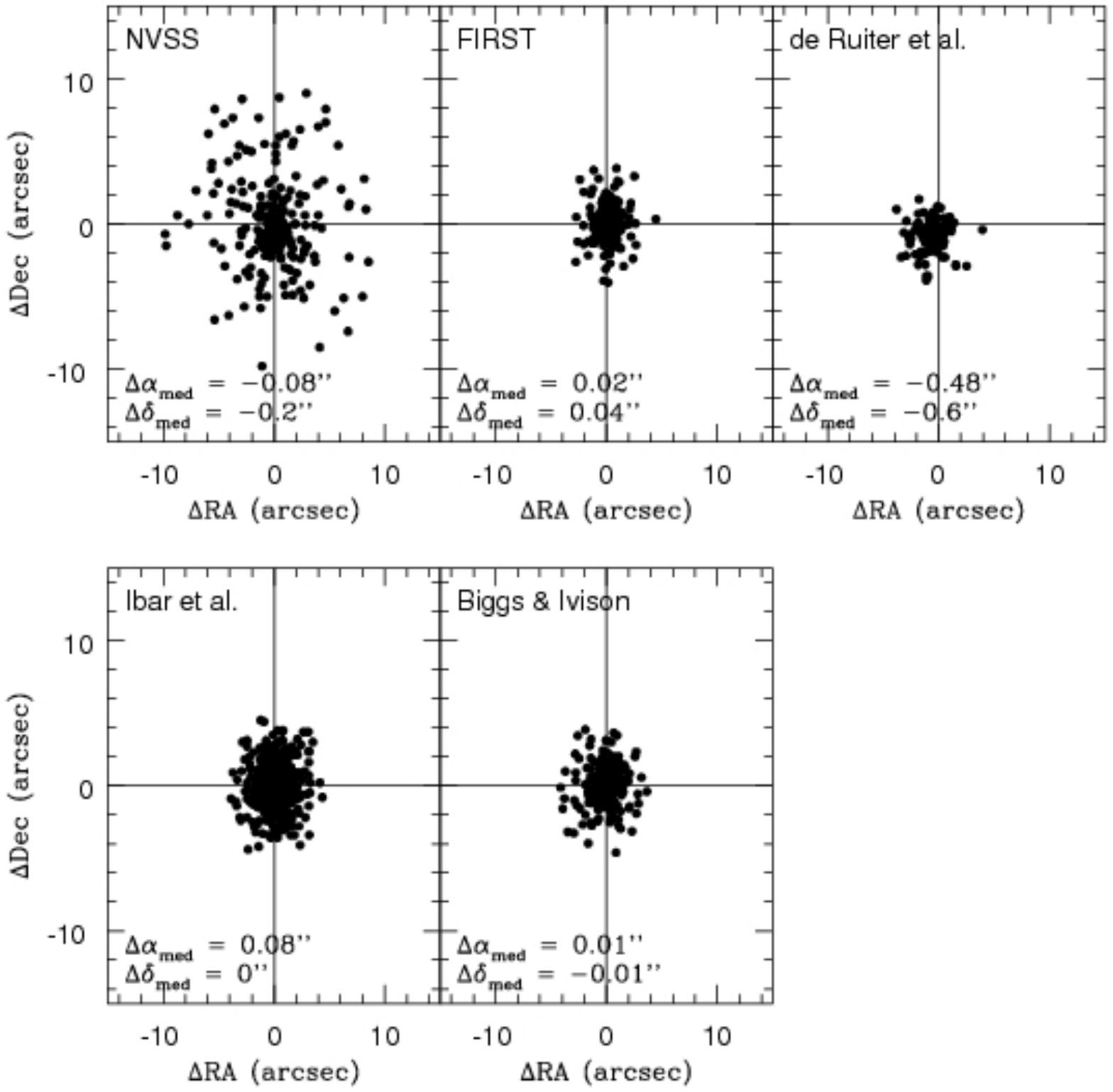}}
\hfill
\begin{minipage}[t]{55mm}
\vspace{-5cm}
\caption{Comparison of source positions between our sample and other existing overlapping catalogues. Median values for $\Delta$RA and $\Delta$Dec are reported in each panel.}
         \label{fig-DXDY}
\end{minipage}
\end{figure*}

The results of this comparison are shown in Figures~\ref{fig-confflux} and \ref{fig-DXDY}. In Figure~\ref{fig-confflux} we have plotted the WSRT against the other catalogue flux densities for all common sources. The plot shows that, despite the different intrinsic resolution of the various surveys (as indicated in the plot panels), the WSRT flux scale is in very good agreement with the NVSS, FIRST, \citet{deruiter97} and \citet{Ibar09} ones over the entire flux range probed by the different samples. This allows us to conclude that our flux calibration errors are within a few percent, in line with expectations, and that no systematic effects have been introduced in the image deconvolution process. On the other hand, the comparison with \citet{Biggs06} shows that our sample is characterized by systematically higher flux densities. This may be partly  due to the significantly different resolution of the two catalogues 
($\sim 10^{\prime\prime}$ against 1.3$^{\prime\prime}$). A similar trend was found by \citet{Ibar09} when comparing their catalogue with the \citeauthor{Biggs06} one. From a detailed analysis of the two samples \citeauthor{Ibar09} however concluded that the systematic differences in flux measurements were to be ascribed to the different approaches used for the source extraction. In particular, \citet{Biggs06} used a fixed beam size to fit a Gaussian to sources which were assigned areas smaller than the beam by the initial extraction procedure. This inevitably yields lower flux measurements (see \citealt{Ibar09}  for more details).

 A similar comparison was repeated for source positions and the result is plotted in Figure~\ref{fig-DXDY}.  Again, the source positions derived for our source catalogue are in very good agreement with those derived for the comparison samples obtained at the VLA and with different calibration strategies. Systematic offsets, if present, can be considered negligible ($\pm 0.01$--$0.1^{\prime\prime}$) with respect to the intrinsic position measurement errors of both our (see Eqs. (\ref{eq-errra}) and (\ref{eq-errdec})) and comparison catalogues. The only exception is represented by the \citet{deruiter97} sample, where larger systematic offsets ($\Delta\alpha$=$-0.5^{\prime\prime}$; $\Delta\delta$=$-0.6^{\prime\prime}$) are found.  However such systematic errors are more likely to be ascribed to the \citeauthor{deruiter97} catalogue, since no significant trend is found in the comparison with the other samples. In summary we can conclude that our source position calibration strategy (through the use of a VLBA secondary calibrator) was successful, and that our reduction strategy has not introduced significant systematic offsets. This is of particular relevance for cross-identification purposes (with other radio and/or optical/IR catalogues), which is an obvious subsequent step for a full scientific exploitation of our sample.

\section{Source Sizes and Resolution Bias}
\label{sec-sizes}

\begin{figure}
   \centering
   \resizebox{9cm}{!}{\includegraphics[trim=0 0 0 150]{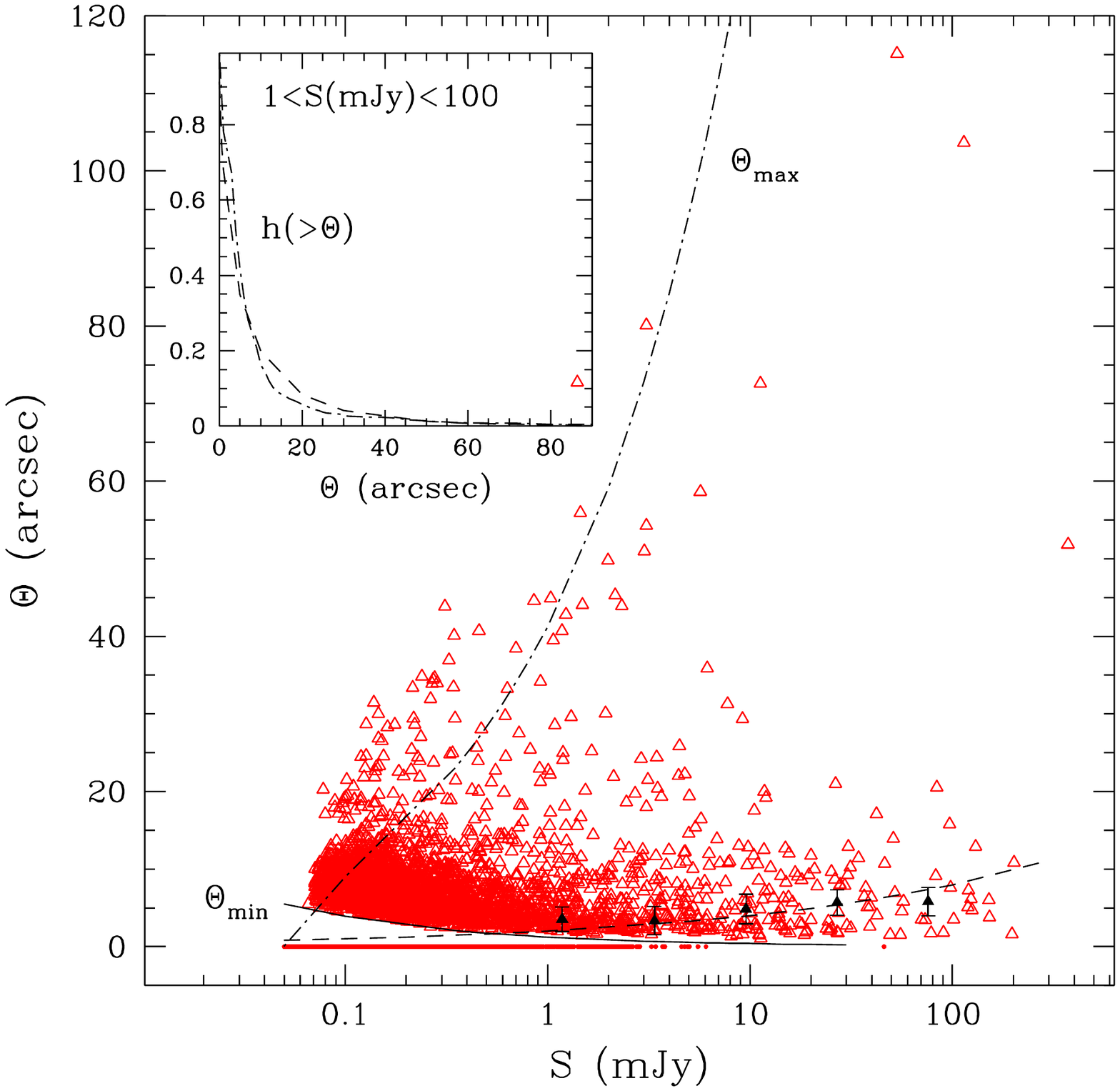}}
   \vspace{-2cm}
      \caption{Deconvolved angular size ($\Theta$=$\Theta_{maj}$) as a function of integrated flux density (for unresolved sources we assume $S_{tot}$=$S_{peak}$). The dot-dashed line represents the size ($\Theta_{max}$), above which the sample becomes incomplete, due to the resolution bias. The solid line indicates the minimum angular size ($\Theta_{min}$),  below which deconvolution is not considered meaningful. The dashed line indicates the median source size  as a function of flux as expected from \citetalias{Windhorst90}) relation. Such line has to be compared to the black points which represent the median source sizes for different flux intervals, as measured in our sample. The inner panel shows the angular size distribution (h($>\Theta$)) proposed by  \citetalias{Windhorst90}; dashed line), compared to the size distribution of sources in our sample (broken solid line).}
         \label{fig-sizeflux}
\end{figure}

Figure~\ref{fig-sizeflux} shows the source Gaussian deconvolved angular sizes as a function of flux density for our sample. The solid line in Figure~\ref{fig-sizeflux} indicates the minimum angular size, $\Theta_{min}$, below which sources are considered point-like, as derived from Eqs.~\ref{eq-sratio} and~\ref{eq-ratio} (see Sect.~\ref{sec-deconv} for more details). In general we can successfully deconvolve $\sim$60\% of the sources in our sample, $\sim$80\% of the sources  with S $\geq$ 0.7 mJy and $\sim$90\% of the sources with S $\geq$ 2.5 mJy. Above such flux limits, where we have a limited number of upper limits, we can reliably undertake a statistical analysis of the source size properties. To this extent, we compare the median angular size measured in different flux intervals for the sources with S $\geq$ 0.7 mJy (black points) and the angular size integral distribution derived for the sources with 1$<$S(mJy)$<$100 (broken solid line in the inner panel) to the ones obtained from the \citet{Windhorst90} relations proposed for deep 1.4 GHz samples: $\Theta_{med}$ = 2$^{\prime\prime}\times $(S$_{\rm{1.4 GHz}}$)$^{0.30}$ (S in mJy) and h($>\Theta$)=exp[-ln 2 ($\Theta$/$\Theta_{med}$)$^{0.62}$].  We notice that the Windhorst et al. relations are widely recognized to provide good statistical descriptions of source sizes at flux densities $\ga 1$ mJy, i.e. at the flux levels probed by our analysis. Indeed our determinations show a very good agreement with the ones of \citetalias{Windhorst90}; see dashed lines in Figure~\ref{fig-sizeflux}). 

We notice that flux losses in extended sources can in principle affect our source parameterization and  cause incompletess in the source catalogue itself. In fact, a resolved source of given $S_{tot}$ will drop below the peak flux density detection threshold more easily than a point source of same $S_{tot}$. This is the so-called \emph{resolution bias}. 
Eq.~\ref{eq-ratio} can be used to give an approximate estimate of the maximum size ($\Theta_{max}$) a source of given $S_{tot}$ can have before dropping below the $S_{peak} $=$5\sigma_{local}$ limit of the source catalogue. Such a limit is represented by the black dot-dashed line plotted in Figure~\ref{fig-sizeflux}. As expected, the angular sizes of the largest sources  approximately follow the estimated $\Theta_{max} - S_{tot}$ relation.

In principle there is a second incompleteness effect, related to the maximum scale at which our WSRT mosaics are sensitive due to the lack of baselines shorter than 36 m. This latter effect can, however, be neglected in our case, because it is smaller than the previous one over the entire flux range spanned by the survey. In fact we expect the sample to become progressively insensitive to source scales larger than 500 arcsec. Moreover, if we assume the angular size distribution proposed by  \citetalias{Windhorst90}) we expect no sources with $\Theta > 500^{\prime\prime}$ in the area and flux range covered by our survey.

\section{Source Counts at 1.4 GH\lowercase{z}}
\label{sec-counts}

  We start by limiting the source count derivation to the mosaic region with local rms noise $< 330$ $\mu$Jy (see discussion in Sect.~\ref{sec-extraction}), corresponding to a total area of $\sim 6$ square degrees. We used all sources brighter than 70 $\mu$Jy (corresponding to $\ga 6\sigma$ in the deepest part of the mosaic) to derive the differential source counts as a function of flux density. This minimizes flux boosting and incompleteness issues at the $5\sigma$ catalogue extraction threshold (see Sect.~\ref{sec-extraction}). Integrated flux densities were used for extended sources and peak flux densities for point-like sources. Each source has been weighted by the reciprocal of its visibility area (A($>S_{peak}$)/$A_{tot}$), as derived from Fig.~\ref{fig-noise} (right panel), by setting $S_{peak}>5\sigma_{local}$.    This is the fraction of the total area over which the source could be detected. 

\begin{figure*}
   \centering
   \resizebox{\hsize}{!}{\includegraphics[angle=-90]{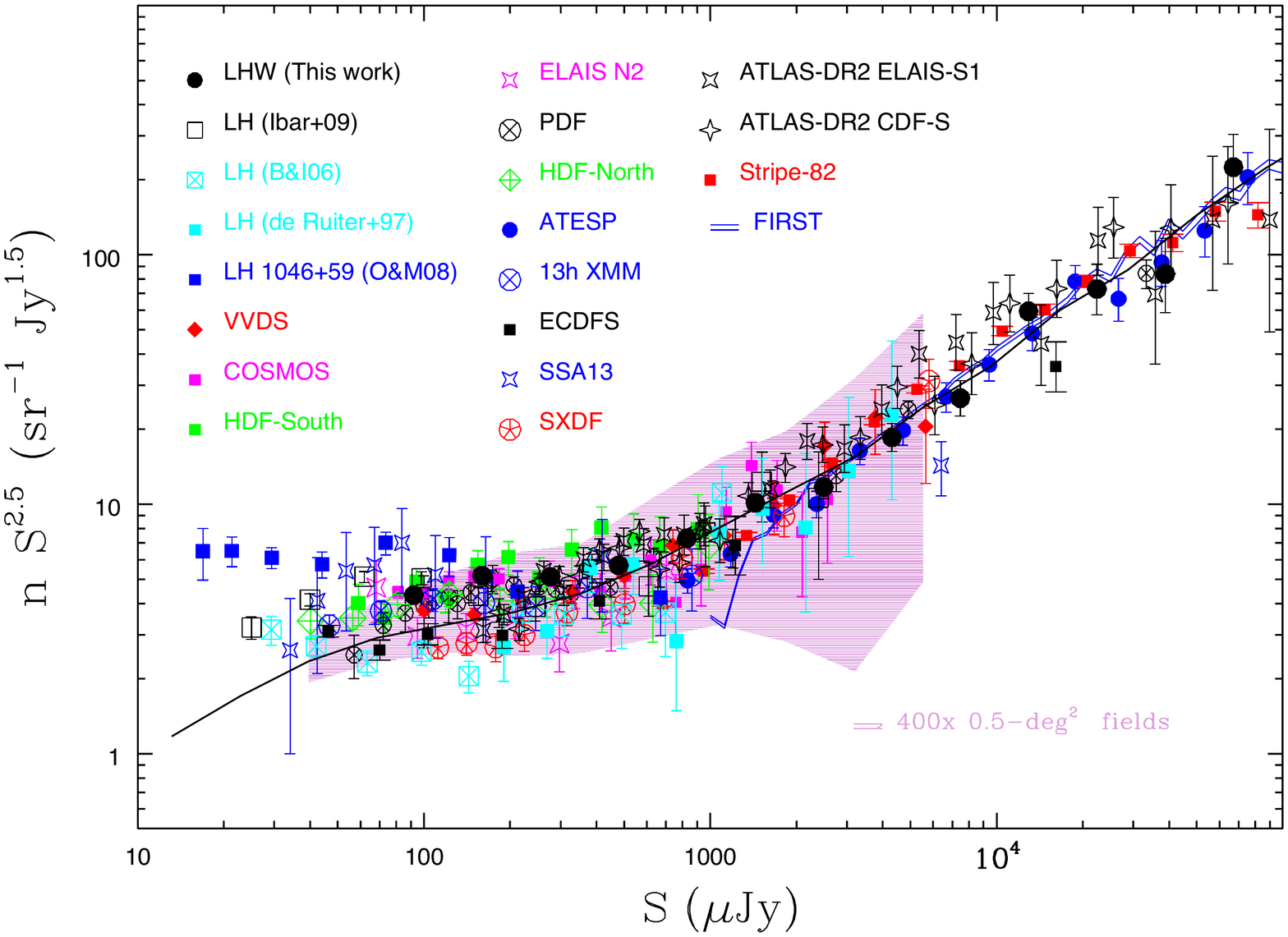}}
 \vspace{-1cm}
      \caption{Normalized 1.4 GHz differential source counts for different samples (as indicated in the figure and in the text). The counts derived from the catalogue  discussed in this work are represented as filled black circles. Vertical bars represent Poissonian errors on the normalized counts. Also shown are the source counts derived from 200 sq. degr. of the semi-empirical simulation of \citet[][S3-SEX, solid line]{Wilman08}, which represent the summed contribution of the modeling of various source populations (RL and RQ AGNs; SFGs), and the predicted spread due to cosmic variance for 0.5 deg$^2$ fields (pink shaded area). They have been  obtained by splitting the S3-SEX simulation in 400 0.5-deg$^2$ fields.}
         \label{fig-diffcounts}
\end{figure*}  
   
 \begin{figure}
   \centering
 \resizebox{\hsize}{!}{\includegraphics[]{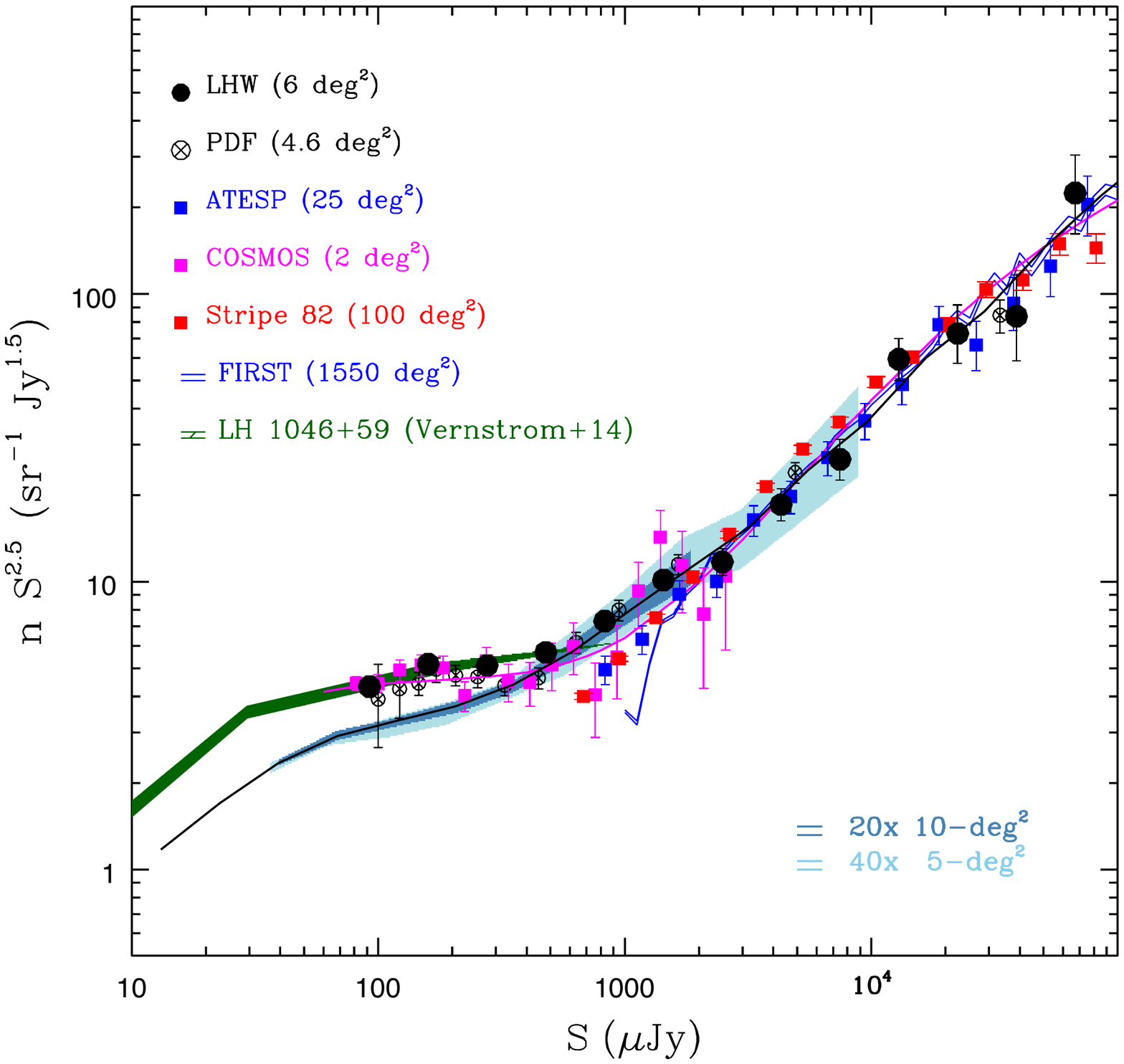}}
 \vspace{-3cm}
      \caption{Normalized 1.4 GHz differential source counts from large-scale surveys  (as indicated in the figure). Vertical bars represent Poissonian errors on the normalized counts. Also shown is the result of the P(D) analysis performed by \citet[][dark green shaded area]{Vernstrom14}. In this case the counts are rescaled from 3 to 1.4 GHz by assuming $\alpha$=-0.7. The solid lines represent the predicted counts from 200 sq. degr. of the S3-SEX simulations   \citep{Wilman08}, while the magenta solid line represents the fit obtained by \citet{Bondi08} from the COSMOS and FIRST source counts. The light and dark blue shaded areas illustrate the predicted cosmic variance effects for survey coverages of 5 and 10 sq. degr. respectively. They have been  obtained by splitting the  S3-SEX simulation in forty 5-deg$^2$ and twenty 10-deg$^2$ fields respectively. }
         \label{fig-widecounts}
\end{figure}  

 \begin{figure}
   \centering
 \resizebox{\hsize}{!}{\includegraphics[]{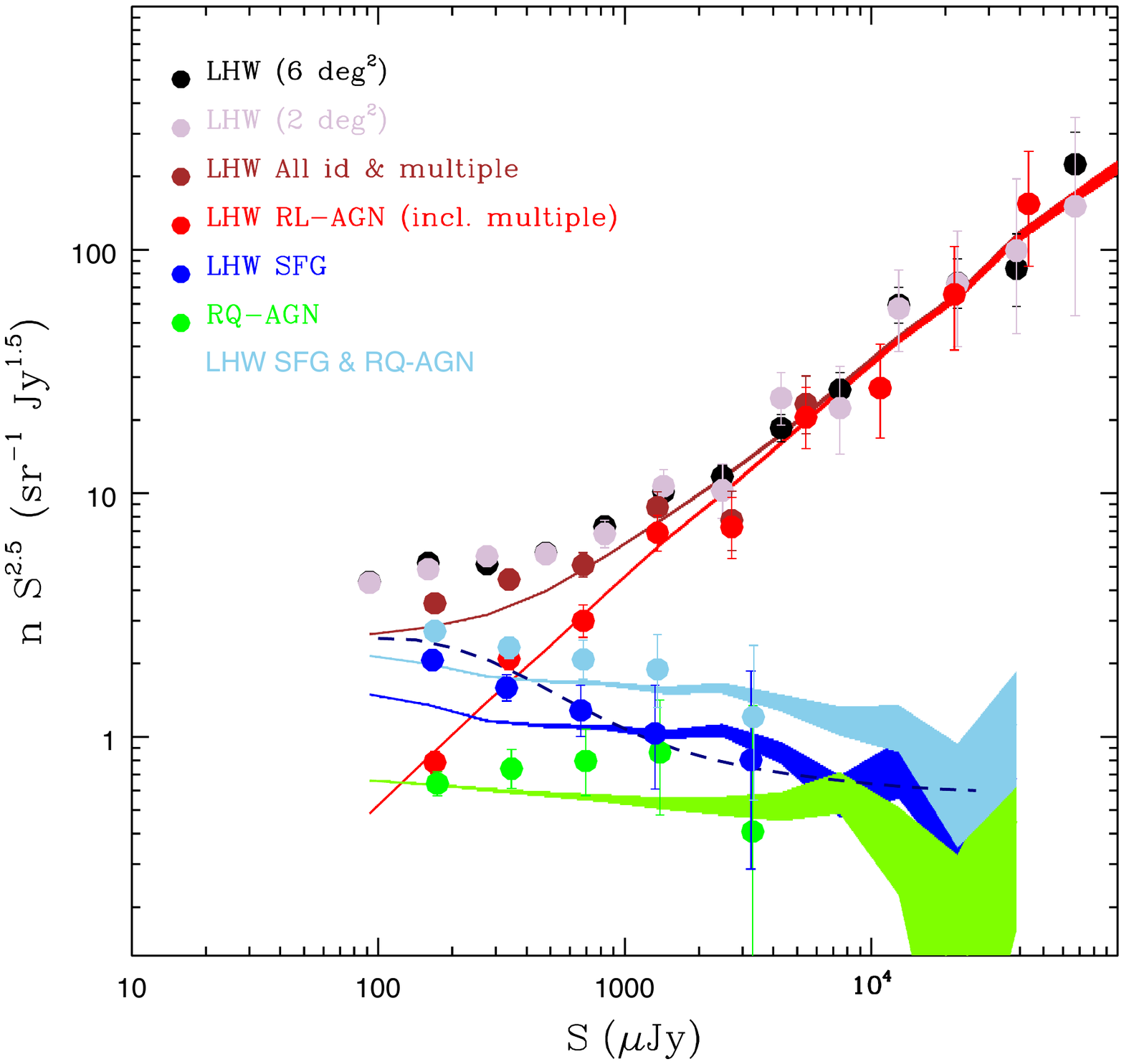}}
 \vspace{-3.cm}
      \caption{Normalized 1.4 GHz differential source counts obtained in the inner $\sim 2$ deg$^2$ region of our mosaic  split in sub-components (filled points) compared to the S3-SEX predictions for a 200 sq. degr. region (solid lines/bands): All radio sources (grey); identified radio sources, including unidentified multi-component sources (brown); RL AGNs, including unidentified multi-component sources (red); SFGs (blue); RQ AGNs (green);
RQ population as a whole (obtained by summing SFGs and RQ AGNs, light blue). 
For a meaningful comparison we applied the same magnitude cut ($11<K_{Vega} <21.5$) at both our identified radio sources and to the S3-SEX catalogues. For the SFG population an additional comparison model is shown (dark-blue dashed line; \citealt{Mancuso17}). The counts obtained in the inner $\sim 2$ deg$^2$ region (grey filled points) are fully consistent with those obtained from the full mosaic (6 deg$^2$, black filled points). Vertical bars represent Poissonian errors on the normalized counts. } 
         \label{fig-countsid}
\end{figure}  

Moreover, we have taken  into account both the catalogue contamination introduced by artefacts (see discussion in Sect.~\ref{sec-artefacts}), and the catalogue  incompleteness, due to the resolution bias discussed in the previous section. The correction $c$ for the resolution bias has been defined following \citet{Prandoni01} as: 
\begin{equation}
      c$=$1$/[$1-h$($>\Theta_{lim}$)$$]$
   \end{equation}
where h($>\Theta_{lim}$) is the integral angular size distribution proposed by \citetalias{Windhorst90}) for 1.4 GHz samples, which turned out to be a good representation of the  source sizes at least down to S$\sim$0.7 mJy (see Sect.~\ref{sec-sizes}). $\Theta_{lim}$ represents the angular size upper limit, above which we expect to be incomplete. This is defined as a function of the integrated source flux density as (see \citealt{Prandoni01}):
\begin{equation}
      \Theta_{lim}$=$max$[$\Theta_{min},\Theta_{max}$]$
   \end{equation}
where $\Theta_{min}$ and $\Theta_{max}$ are the parameters defined in Sect. 8. The $\Theta_{min} - S$ relation (solid line in Fig.~\ref{fig-sizeflux}) is important at low flux levels where $\Theta_{max}$ (black dot-dashed line in Fig.~\ref{fig-sizeflux}) becomes unphysical (i.e. $\rightarrow$0). In other words, introducing $\Theta_{min}$ in the equation takes into account the effect of having a finite synthesized beam size (that is $\Theta_{lim}\gg0$ at the survey limit) and a deconvolution efficiency which varies with the source peak flux. 

\begin{table}
\centering
      \caption{1.4 GHz source counts as derived from our survey. The format is the following: {\it Column 1:} flux interval ($\Delta$S); {\it Column 2:}  geometric mean of the flux density ($\langle S\rangle$); {\it Column 3:}  number of sources detected ($N_S$); {\it Column 4:}  differential counts normalized to a non evolving Euclidean model (n $S^{2.5}$). Also listed are the Poissonian errors (calculated following \citealt{Regener51}) associated to the normalized counts. }
         \label{tab-counts}
\begingroup

\renewcommand{\arraystretch}{1.5} 
      \begin{tabular}{|c|c|c|r|}
      \hline
      $\Delta S$ & $\langle S\rangle$ & $N_{S}$ & $d$N/$d$S $S^{2.5} \pm \sigma$ \\
      (mJy) & (mJy) &  & ($sr^{-1} Jy^{1.5}$) \\
      \hline 
     0.070 -  0.121  &   0.092 & 1896   &  4.32$^{\text{+}0.10}_{-0.10}$ \\
     0.121 -  0.210  &   0.160 & 1462   &  5.17$^{\text{+}0.14}_{-0.14} $ \\ 
     0.210 -  0.364  &   0.276  & 801   &     5.12$^{\text{+}0.18}_{-0.19} $   \\
     0.364 -  0.630  &    0.479 &  457   &  5.69$^{\text{+}0.27}_{-0.28}$   \\    
     0.630 -  1.091  &   0.829 &  282   &   7.30$^{\text{+}0.43}_{-0.46} $  \\      
     1.091 -  1.890  &   1.436 &  181 &   10.14$^{\text{+}0.75}_{-0.81} $  \\      
     1.890 -  3.274  &   2.487 &  100 &    11.7$^{\text{+}1.2}_{-1.3}$    \\
     3.274 -  5.670  &   4.308 &   67  &  18.5$^{\text{+}2.3}_{-2.5}$  \\  
     5.670 -  9.821  &   7.462 &   42 &   26.6$^{\text{+}4.1}_{-4.7}$  \\     
     9.821 - 17.01   &  12.92 &   41  &  59.4$^{\text{+}9.3}_{-10.7}$  \\       
    17.01 -   29.46   & 22.39 &   22  &  73$^{\text{+}16}_{-19}$  \\    
    29.46 -   51.03  &   38.77 &   11  &  84$^{\text{+}25}_{-33}$  \\       
    51.03 - 88.39   &  67.16 &   13 &   224$^{\text{+}62}_{-79}$  \\       
    88.39 - 153.1 &   116.3 &   10 &  391$^{\text{+}124}_{-163}$  \\   
   153.1 -  265.2   & 201.5  &  5   & 356$^{\text{+}159}_{-230}$  \\   
         \hline
      \end{tabular}
      \endgroup
      \end{table}

   The differential source counts normalized to a non-evolving Euclidean model (n $S^{2.5}$) are  listed in Table~\ref{tab-counts} and shown in Figure~\ref{fig-diffcounts} (filled black circles).  Our source counts are compared with others available at 1.4 GHz from the literature, either  in the LH region \citep{deruiter97, Biggs06,Owen08,Ibar09} or in other regions of the sky. 
  This includes all known deep fields, from single pointings like 13h XMM \citep{Seymour04}, HDF South  and North \citep{Huynh05,Biggs06}, ELAIS N2 \citep{Biggs06}, SXDF \citep{Simpson06}, SSA13 \citep{Fomalont06}  and ECDFS \citep{Padovani15}, to wider-area ($>$1 sq. degr.) regions, like PDF \citep{Hopkins03}, VLA-VVDS \citep{Bondi03}, VLA-COSMOS \citep{Bondi08}, ATLAS \citep{Hales14}, to shallower large ($\gg$10 sq. degr.) surveys like  ATESP \citep{Prandoni01}, SDSS Stripe 82 \citep{Heywood16} and FIRST \citep{White97}. 
Also shown are the source counts derived from the semi-empirical sky simulation developed in the framework of the SKA Simluated Skies project (S3-SEX, \citealp{Wilman08,Wilman10}; solid line), which represent the summed contribution of the modeling of various source populations (RL and RQ AGNs; SFGs).

Figure~\ref{fig-diffcounts}  illustrates very well the long-standing issue of the large scatter (exceeding Poisson fluctuations) present at flux densities $\la 1$ mJy.  The main causes for this considerable scatter can be either cosmic variance (source clustering) or survey systematics introduced by e.g. calibration, deconvolution and source extraction algorithms, or corrections applied to raw data to derive the source counts. These issues have been extensively discussed in the recent literature. \citet{Heywood13} compared the observed source counts with samples of matching areas extracted from the S3-SEX simulations \citep{Wilman08,Wilman10}, that include a recipe for source clustering, and concluded that the observed scatter is dominated by cosmic variance, at least down to 100 $\mu$Jy. This is clearly illustrated by the pink shaded area shown in Fig.~\ref{fig-diffcounts}, showing the predicted source counts' spread due to cosmic variance for  typical areas of deep radio fields. This has been obtained by splitting the  S3-SEX simulation in 400 0.5-deg$^2$ fields.  

There are however a few exceptions. The most notably one is the anomalously high number counts estimate obtained by \citet{Owen08} in the LH 1046+59 field (filled blue squares in Fig.~\ref{fig-diffcounts}).  New confusion-limited, lower resolution VLA observations of the same field obtained at 3 GHz demonstrated  that this is the result of an over-estimated resolution bias correction \citep{Condon12,Vernstrom14}. Another exception could be  the counts' estimate obtained in the LH region by \citet[][cyan squared crosses]{Biggs06}, which tends to be  low (even if still consistent with cosmic variance). As we demonstrated  in Sect.~\ref{sec-errors}, this sample suffers from flux underestimations. Also in this case the low counts are more likely to be ascribed  to technical problems, rather than mere cosmic variance.  

\section{Constraints from wide-area source counts}
\label{sec-countsid}

A more robust  view of the 1.4 GHz source counts can  be obtained by using the widest-area samples available to date. Cosmic variance depends on a combination of area coverage and depth. For a given allowed cosmic variance, the deeper the survey, the larger the volume sampled, the smaller the area coverage requirement.  Figure~\ref{fig-widecounts} shows the source counts derived from the largest 1.4 GHz surveys obtained so far  and fully sensitive to the flux range  $\sim 0.1 - 1$  mJy (COSMOS, PDF and this work), together with the largest  surveys that probe flux densities $\ga 1$ mJy (ATESP, Stripe 82 and FIRST).  Figure~\ref{fig-widecounts} clearly shows that the scatter below 1 mJy is considerably reduced, when limiting to widest-area surveys, and is mostly consistent with Poisson fluctuations. Our LH survey (counts derived over $\sim 6$ deg$^2$) provides the most robust statistics between $\sim 0.1$ and  1 mJy, while  the FIRST (1550 deg$^2$) provides excellent estimates  above a few mJy. 
Recent wide-area, deep surveys obtained at frequencies different from 1.4 GHz (like e.g. the VLA-COSMOS 3 GHz Large Project - \citealp{Smolcic17} - or the ATCA XXL Survey at 2.1 GHz - \citealp{Butler17}) are not included here to avoid  introducing unwanted uncertainties or systematic effects.  Indeed  the extrapolation of the counts to 1.4 GHz is very sensitive to the assumed spectral index. In addition adopting the same (average or median) spectral index value for all the sources may not be appropriate. There are indeed indications that the spectral index may flatten at mJy/sub-mJy regimes (e.g. \citealp{Prandoni06,Whittam13}) and re-steepen again at $\mu$Jy levels, depending on the flux-dependent AGN/SFG  source mix (e.g. \citealp{Owen09}). The only exception we make is for the analysis performed at 3 GHz in the LH 1046+59 field by \citet{Vernstrom14}, that provides the most  reliable constraint available so far down to flux densities $\la 10$ $\mu$Jy. This counts determination is obtained from the so-called P(D) analysis, which allows to investigate the source statistical properties well below the confusion limit of the survey. As such, the P(D) analysis tends to be less prone to resolution effects, as well as to  cosmic variance (as it probes larger volumes), and can provide more reliable results even in case of limited survey areas. The  source counts derived from wide-area sub-mJy surveys  are fully consistent  with the results of the P(D) analysis of \citet{Vernstrom14}.

The only remaining uncertainties are around the knee of the distribution ($\sim 1 $ mJy), where the shallower surveys seem to suffer from incompleteness, and some discrepancy is found between the various surveys. 

The measured source counts tend to be  higher than the predicted model from \citet[][black solid line]{Wilman08,Wilman10} in the flux range 10--400 $\mu$Jy. This discrepancy appear statistically significant and cannot be accounted for by mere cosmic variance, whose effects on 5 and 10 deg$^2$ survey scales are illustrated by the light and dark blue shaded regions respectively.  The P(D) analysis of \citet{Vernstrom14} suggests that the observed excess may be present down to very faint flux densities.

In order to assess which component of the faint radio population is most likely responsible for the measured excess observed in the  source counts below a few hundreds $\mu$Jy, we exploited the multi-wavelength information available in the LH region. This allowed us to get a first characterization of  the radio sources extracted from our mosaic. 
The identification and classification process will be fully discussed in a forthcoming paper, where the multi-band properties of the faint radio population in the LH field will be presented. Here we only summarize in broad terms the method followed and the diagnostics used. 

In order to identify the counterparts of the WSRT 1.4 GHz sources we used the so-called {\it SERVS Data Fusion}  \citep{Vaccari10,Vaccari15}, a mid-infrared-selected catalog combining the far-ultraviolet-to-sub-millimeter datasets described in Sect.~\ref{sec-multiband}, as well as the available photometric and spectroscopic redshifts. 
Since we are interested to probe the counts' sub-mJy regime  we limited our analysis to the inner $\sim 2$ deg$^2$ of our mosaic, where a roughly uniform  rms noise of  $\sim 11$ $\mu$Jy is measured (see Sect.~\ref{sec-noise}), and which is fully covered by the UKIDSS-DXS deep K-band mosaic \citep{Lawrence07}. In addition only radio sources brighter than 120 $\mu$Jy (i.e. with SNR $>10\sigma$) were considered. This was motivated by the relatively poor resolution of our WSRT observations ($\sim 10$ arcsec) which results in positional errors of the order of $\sim 1$ arcsec at the faintest ($5\sigma$) limit of the radio catalogue (see Sect.~\ref{sec-errors}). This, combined with the high number density of the confusion-limited SERVS dataset, that prevents us to push our identification search beyond 1.5-2 arcsec (to keep contamination under control), means that our identification procedure gets progressively incomplete going to lower flux densities. At 120 $\mu$Jy we estimate the incompleteness to be $\sim 10-15\%$.  In summary we restricted our sample to 1110 single-component radio sources, 80\% of which were identified. For the reasons outlined above we did not attempt a matching of the multi-component radio sources, as the error associated to their position is even larger  (typically of the order of few arcseconds). Nevertheless we know that multi-component sources are radio galaxies. We can therefore directly assign these latter sources (45 with $S>120$ $\mu$Jy) to the RL AGN class. 

The identified single-component radio sources were classified using multi-band diagnostics, as described below. We first identified the RL AGN component, and then proceeded with the separation of SFGs from AGNs in the remaining RQ population. RL AGNs were primarily identified through their radio excess. In particular we used  the well-known method based on the observed  24 $\mu$m to 1.4 GHz flux density ratio 
($q_{24 \,  obs} $=  log[$S_{{\rm 24 \,\mu m}}$/$S_{{\rm 1.4 \, GHz}}$]; see \citealp{Bonzini13}). For each source the  $q_{24 \, obs}$  parameter is compared to  the one expected for SFGs as a function of redshift  (as illustrated in Fig.~\ref{fig-24mic}). Sources not detected at  24 $\mu$m, and characterized by 24 $\mu$m to 1.4 GHz flux density ratio upper limits not stringent enough for  a reliable classification, were classified based on their  red IRAC colors (i.e. their 8.0 to 4.5 $\mu$m and 5.8 to 3.6 flux density ratios; see \citealp{Luchsinger15} for more details). 
In the recent literature RQ AGNs in deep radio-selected samples are separated from SFGs based on their  IRAC colors  and, when available, X-ray luminosities (see e.g. \citealp{Bonzini13}). In the LH region  the X-ray band information is limited to the two deep fields observed by XMM and Chandra (see Fig.~\ref{fig-mosaic}), and both are located at the periphery of our  1.4 GHz mosaic. We therefore decided to complement the IRAC-color-based classification (\citealp{Lacy04,Lacy07,Stern05,Donley12};  see example in Fig.~\ref{fig-IRAC}) with other diagnostic diagrams discussed in the literature, that combine IRAC with Herschel or  K-band information (e.g. \citealp{Kirkpatrick12,Messias12}).  The latter  (shown in Fig.~\ref{fig-KI}) proved to be particularly useful, thanks to the high detection rate at K-band of our identified sources ($\sim 90\%$). The above procedures allowed us to classify 99\% of the identified sources. 

The results in terms of source counts are illustrated in Fig.~\ref{fig-countsid}, where we compare the counts obtained from the identified sources in our sample (split in several sub-components) to  those expected from the S3-SEX simulations, after applying the same cut  in magnitude to both our identified sample and the S3-SEX simulations ($11<K_{Vega} < 21.5$). RL AGNs (red points) nicely follow the S3-SEX predictions (red solid line) over the entire flux range probed by our sample. The RQ population (i.e. the sum of SFGs and RQ AGNs), on the other hand, show an excess with respect to the predicted counts below $\sim 500$ $\mu$Jy (see  light blue points and solid line). The observed excess is marginal, but affected by the aforementioned incompleteness. When splitting the RQ population in its two main components, SFGs (shown in blue) and RQ AGNs (shown in green), the excess becomes more relevant for SFGs. RQ AGNs do show some excess at intermediate fluxes, but given the large error bars, it cannot be considered statistically significant.  We caveat, though, that the selection criteria used to identify RQ AGN (essentially based on IR and K-band information) will miss objects that present  AGN signatures only in the X-ray band or in their optical spectra. Therefore we cannot exclude that some of the excess over the S3-SEX models, now entirely attributed to SFGs, might be associated with RQ-AGNs.

The SFG component in the S3-SEX simulations has been modeled starting from the  well-constrained 1.4 GHz local luminosity function of SFGs (see e.g. \citealp{Yun01}) and assuming a pure luminosity evolution of the form (1+z)$^{3.1}$ (for both normal and starburst galaxies) out to z=1.5,  with no further evolution thereafter. This evolutionary form can reproduce  the observed number density in the local ($z<0.3$) universe (see e.g. \citealp{Mauch07}), that dominate at flux densities $\ga 2$ mJy, and is consistent with other constraints obtained by combining the observed global star formation rate (SFR) density evolution and the 1.4 GHz radio source counts \citep{Hopkins04}.  Nevertheless  our measured source counts suggest a somewhat steeper evolution for SFGs going to  lower flux densities (or higher redshifts).   

A full discussion of the evolutionary properties of radio-selected star-forming galaxies is beyond the scopes of this paper, but it is worth mentioning that a steeper evolution of SFGs  is supported by the work of \citet{Mancuso16,Mancuso17}, who proposed a novel model-independent approach, where they combine UV and far-IR data to trace the evolution of the intrinsic SFR function over the entire redshift range $z\sim 0-10$ (see \citealp{Mancuso16} for details). Their results are claimed to be less prone to dust extinction effects especially at high redshifts  ($z > 3$) and  at large star formation rates ($SFR\ga 100$ M$_{\odot}$ yr$^{-1}$), where they predict a  heavily dust-obscured galaxy population, in excess to what found by previous works.  By using standard prescriptions to convert SFRs into 1.4 GHz luminosities, \citet{Mancuso17} were able to get a novel prediction of the contribution of SFGs to 1.4 GHz source counts. 
Interestingly, \citet{Mancuso17} prediction (dark blue dashed line) seems to better reproduce the shape of the observed  SFG counts, providing a good match to both the excess at low flux densities (especially when considering the 10-20\% incompleteness of the sample), and the observations at brighter fluxes (progressively dominated by local radio sources). 

For completeness it is worth mentioning that other recent works (mainly based on the VLA-COSMOS 3 GHz Large Project; \citealp{Smolcic17}) have proposed different models for radio-selected SFGs. For instance \citet{Novak17} found that their data are consistent with a pure luminosity evolution of the form (1+z)$^{3.16 \pm 0.2}$ out to the redshift limit of the survey (i.e. $z\sim 4-5$). In addition they report  a negative evolution of the IR/radio correlation for SFGs. This evolution takes the form of  $\sim$(1+z)$^{\alpha}$, with $\alpha$= $-0.14\pm 0.01$. 
\citet{Bonato17} claim that any evolution of the IR/radio correlation should be mild ($|\alpha| \la 0.16$), as larger values would produce total source counts, inconsistent with the ones observed. 


\section{Summary and Future Perspectives}
\label{sec-summary}

In this paper, part of the Lockman Hole Project, we have presented a deep 1.4~GHz mosaic of the Lockman Hole area obtained using the WSRT. 
The final image  covers an area of 6.6 square degrees, has a resolution of 11 x 9  arcsec and reaches an rms of 11 $\mu$Jy beam$^{-1}$ at the centre of the field. This is the widest-area survey reaching such sensitivity so far.
The catalogue extracted includes 6000 sources and it has already been exploited in a number of studies. For instance, \citet{Mahony16} used it in combination with LOFAR 150 MHz data and other public catalogues for a multi-frequency radio spectral analysis of the mJy radio population in the Lockman Hole region. \citet{Brienza17} expanded the radio spectral study to remnant radio AGNs in the same region.

The obtained source counts provide very robust statistics in the flux range $0.1 < S < 1$ mJy, and are in excellent agreement with other robust determinations obtained at lower and higher flux densities. In particular, when considering only the source counts from wide area surveys the source counts are very well constrained down to $\ga 10$ $\mu$Jy. This allows us to start using source counts to constrain source evolutionary models; something that was prevented so far, due to the long-standing issue of the large scatter present at flux densities $\leq 1$ mJy, and mainly motivated by cosmic variance \citepalias{Heywood13}.  
Notably, the overall wide-area source counts show a clear excess with respect to the counts predicted by the semi-empirical radio sky simulations developed in the framework of the SKA Simulated Skies project \citep{Wilman08,Wilman10}, in the flux range 10--400 $\mu$Jy. This discrepancy appears statistically significant and cannot be accounted for by cosmic variance. 

Making use of the multi-wavelength information available for the LH area, we have separated the sources in RL-AGN, RQ-AGN and SFG and derived the source counts for these three separate groups. A preliminary analysis of the identified (and classified) sources suggests this excess is to be ascribed to star forming galaxies, which seem to show a steeper evolution than predicted. The counts for RL and RQ AGN, on the other hand,  appears to be in line with established models, like those implemented in the S3-SEX simulations. 
A steeper evolution of SFGs is supported by other recent observational work (e.g. \citealp{Novak17}), and in particular by the novel model-independent approach proposed by \citet{Mancuso16,Mancuso17}, who combine UV and far-IR data to trace the evolution of the intrinsic SFR function over the redshift range $z\sim 0-10$. A detailed and more complete analysis of the evolutionary properties of the LH sources, in comparison with other  observational and modeling works will be the subject of a forthcoming paper. In the future we also plan to make use of upgraded multi-band information from the  {\it Herschel Extragalactic Legacy Project} (HELP;  \citealp{Vaccari16}).

The unique combination of sensitivity and area coverage of the the WSRT Lockman Hole mosaic, makes it ideal to be used as a test case for the exploitation of future wider-area surveys obtained by new (or upgraded) radio facilities. Indeed, significantly larger but equally deep images will be obtained using the phased-array-feed system Apertif installed now on the WSRT array \citep{Oosterloo09} as well as ASKAP.
It is worth to notice that the large field-of-view of Apertif will ensure a good match in spatial covering with LOFAR 150~MHz  fields. The sensitivity of the two radio telescopes is also comparable for radio sources with typical spectral index of $-0.7$ (see \citealp{Shimwell17}), opening exciting perspectives for statistical multi-frequency studies of the faint radio sky.

Last but not least, new GHz telescopes like Apertif, ASKAP and MeerKat (see \citealp{Booth12}) offer broad band capability that will allow to obtain simultaneously radio continuum and atomic neutral hydrogen, \HI-21cm, information for the sources in the redshift range sampled by the observing band. Serendipitous \HI\ detections  in famous fields (e.g. FLS), have been already reported using data from traditional radio telescope like the WSRT (see \citealp{Morganti04}) demonstrating the feasibility. A first stacking experiment to study the \HI\ in various classes of radio sources in the Lockman Hole area has also been presented in \citet{Gereb13}. The new generation of radio telescopes will make this a standard procedure expanding the information available for each class of galaxies.

\section*{Acknowledgements}
IP thanks Claudia Mancuso for providing the evolutionary track shown in Figure~\ref{fig-countsid}, which refers to the model discussed in  Mancuso et al. (2017 ). This work has been supported by  financial contribution from PRIN-INAF (2008) and from ASTRON. IP and MV acknowledge support from the Ministry of Foreign Affairs and International Cooperation, Directorate General for the Country Promotion (Bilateral Grant Agreements ZA14GR02 and ZA18GR02).  RM gratefully acknowledges support from the European Research Council under the European Union's Seventh Framework Programme (FP/2007-2013) /ERC Advanced Grant RADIOLIFE-320745. This work is based on research supported by the National Research Foundation of South Africa (Grant Number 113121). The Westerbork Synthesis Radio Telescope is operated by the Netherlands Institute for Radio Astronomy ASTRON, with support of NWO.









\appendix

\section{Source classification}

In this Appendix we show the multi-band diagnostic diagrams discussed in Sect.~\ref{sec-countsid}, where the source classification procedure is presented. The color-coding in the figures refers to the final classification of the sources, based on the information at all available wavebands.

 \begin{figure}
   \centering
  \resizebox{\hsize}{!}{\includegraphics[angle=0]{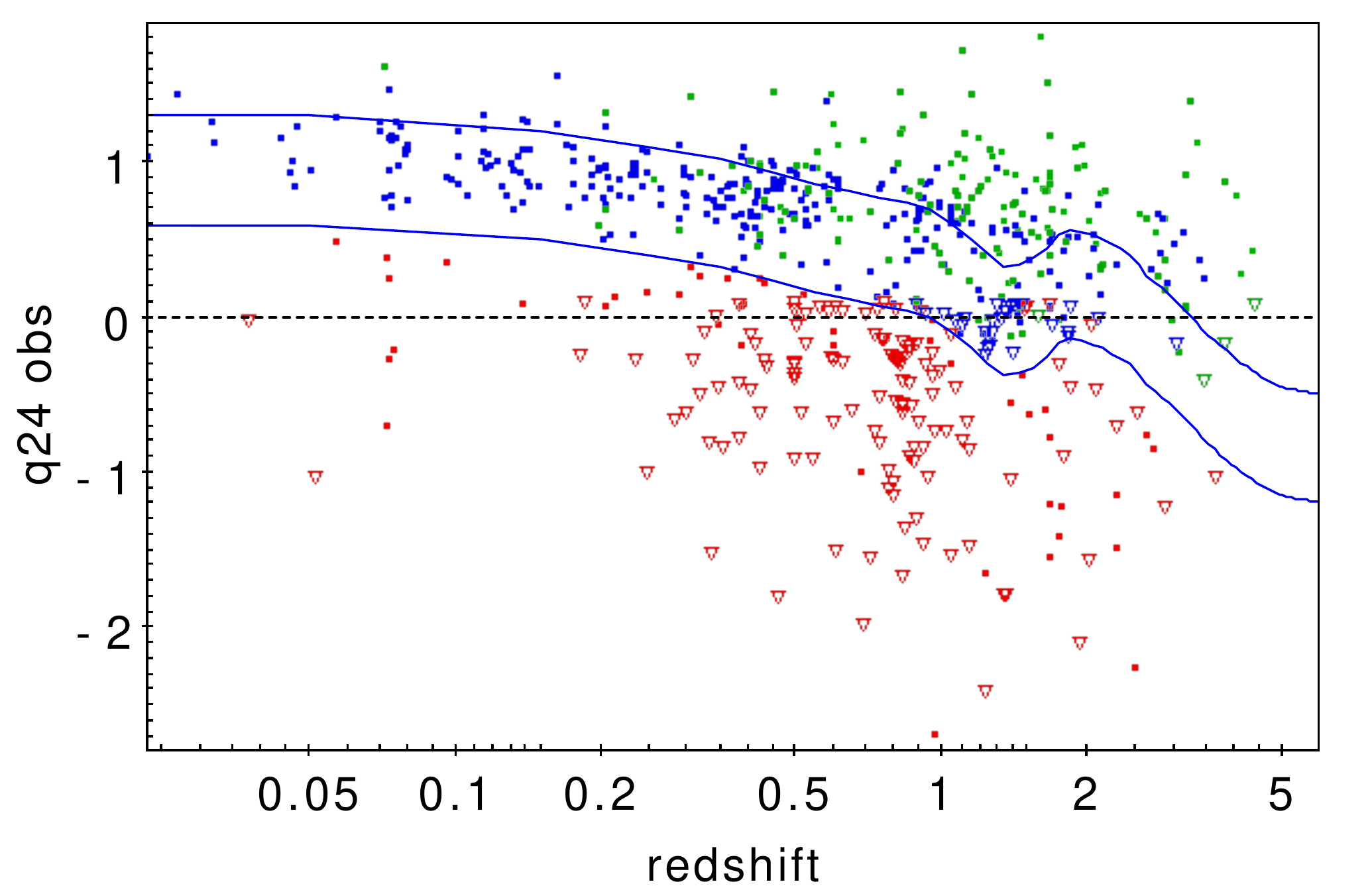}}
      \caption[]{$q_{24 \,  obs} $ values for the identified sources with $S\geq 0.12$ mJy with reliable redshift.  For sources not detected at 24 $\mu$m we set upper limits (downward triangles).  The two solid lines represent the redshifted  M82 template normalized to the local average $q_{24 \, obs}$ value  for star forming galaxies  (upper line), and the threshold adopted to identify  radio loud AGNs (lower line), defined by assuming a  $\geq 2\sigma$  radio excess with respect to the M82 template. The sources which satisfy the radio loudness criterion are shown in red.  The red upper limits above the adopted threshold are sources that have IRAC colors of elliptical galaxies (the typical hosts of radio galaxies) and that we added  to the RL AGN class (see text for more details). The RQ population is divided in SFGs (blue) and RQ AGNs (green) as resulting from the classification procedure discussed in the text. The horizontal dotted line indicate the $q_{24 \, obs}\sim 0.0$ threshold assumed for sources with no redshift available (not plotted).  }
        \label{fig-24mic}
\end{figure}  

\begin{figure}
   \centering
  \resizebox{\hsize}{!}{\includegraphics[angle=0]{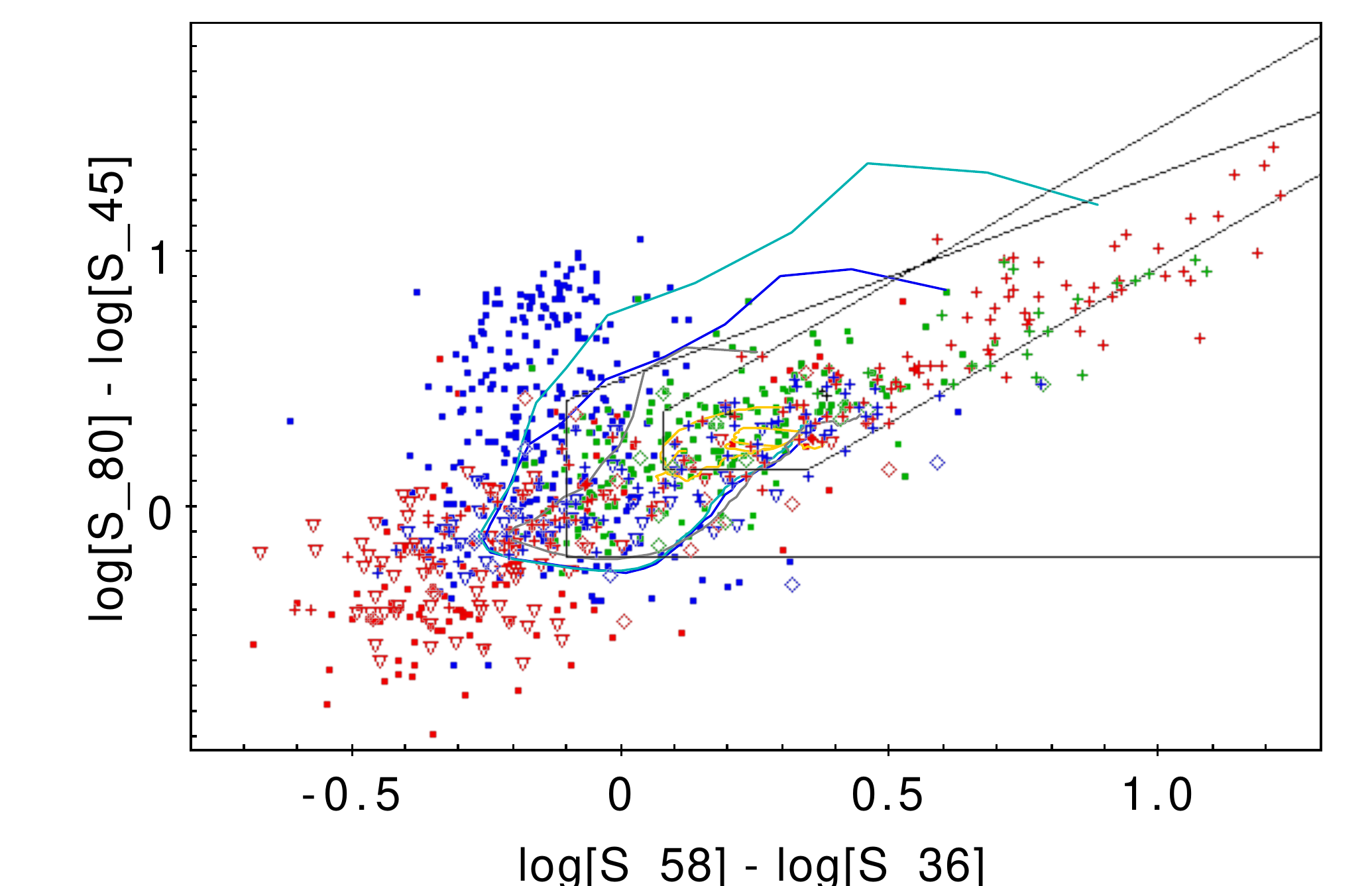}}
      \caption[]{IRAC color - color diagram for the identified sources with $S\geq 0.12$ mJy: $\log{S_{80}} - \log{S_{45}}$ vs. $\log{S_{58}} - \log{S_{36}}$.  Source color-coding is as follows: RQ-AGNs (green); SFGs (blue); RL AGNs (red). Sources detected at all four IRAC bands (3.6, 4.5, 5.8 and 8.0 $\mu$m) are shown as filled points; sources not detected at 8.0 $\mu$m only are shown as downward triangles; sources not detected at 5.8 $\mu$m only are shown as open diamonds; sources not detected at both 5.8 and 8.0 $\mu$m are shown as crosses. Sources detected only at one of the SERVS channels (3.6 or 4.5 $\mu$m) cannot be constrained and are not shown.  The black solid lines represent the AGN wedges as defined by \citet[][larger wedge]{Lacy04,Lacy07} and \citet[][smaller wedge]{Donley12}.   Also  shown are the redshifted SEDs for a number of prototypical classes (from Polletta et al. 2007). Line color-coding is as follows: old (13 Gyr) elliptical galaxy (red); two star-forming galaxy templates: M82 (blue) and Arp 220 (light blue); QSO (green); Seyfert 1 (orange) and Seyfert 2 (gray) galaxies. Redshift increases along the lines anti-clockwise.}
        \label{fig-IRAC}
\end{figure}  

  \begin{figure}
   \centering
  \resizebox{\hsize}{!}{\includegraphics[angle=0]{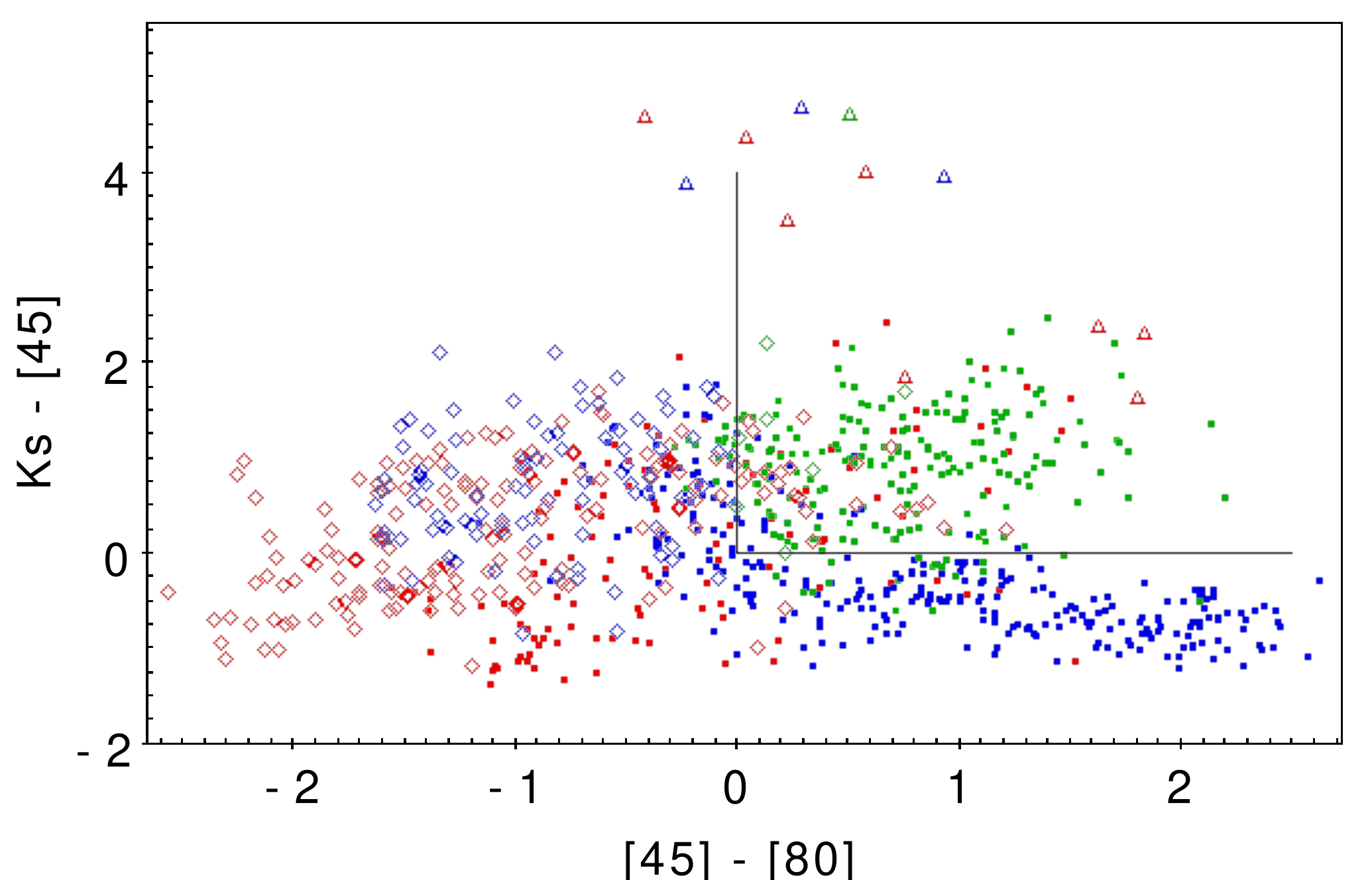}}
      \caption[]{KI diagnostic plot for the identified sources with $S\geq 0.12$ mJy:  4.5 - 8.0 $\mu$m against $K_{s}$ - 4.5 $\mu$m (all quantities are AB magnitudes).  Source color-coding as in Fig.~\ref{fig-IRAC}. Sources detected at all three bands (K-band, 4.5 and 8.0 $\mu$m) are shown as filled points; sources not detected at 8.0 $\mu$m only are shown as open diamonds; sources not detected at K-band only are shown as upward triangles. Sources detected only at 4.5 $\mu$m are shown as crosses. Sources not detected at any of the three bands cannot be constrained and are not shown. The black solid lines represent the AGN wedge as defined by \citet{Messias12}. }
        \label{fig-KI}
\end{figure}



\bsp	
\label{lastpage}
\end{document}